\documentclass[eqsecnum,amsmath,preprintnumbers,nofootinbib,aps]{revtex4} 
\usepackage{tikz}
\usetikzlibrary{calc,trees,positioning,arrows,chains,shapes.geometric,decorations.pathreplacing,decorations.pathmorphing,shapes,matrix,shapes.symbols}

\usepackage{amsmath}
\usepackage{graphicx}
\usepackage{booktabs}
\usepackage{feynmp}
\usepackage{hyperref}
\usepackage{bbold}
\usepackage{eufrak}
\usepackage{color} 
\usepackage{etex}
\reserveinserts{18}
\usepackage{morefloats}

 \usepackage{longtable}

\newcommand{\Slash}[1]{{\ooalign{\hfil/\hfil\crcr$#1$}}} 
\def\beq{\begin{equation}}
\def\eeq{\end{equation}}

\def\bea{\arraycolsep .1em \begin{eqnarray}}
\def\eea{\end{eqnarray}}
\def\Tr{{\rm Tr}}

\def\ga{\gamma}
\def\g{\mathfrak{g}}
\def\eps{\epsilon}

\def\la{\lambda}

\def\sdet{ {\rm sdet} } 
\def\STr{ {\rm STr } }
\def\S{ \mathcal{S}}
\def\C{ \mathcal{C}}
\def\O{ \mathcal{O}}
\def\V{ \mathcal{V}}

\def\n{ {\mathfrak{n}}}
\def\m{ {\mathfrak{m}}}

\def\eq#1{(\ref{#1})}

\def\s0#1#2{\mbox{\small{$ \frac{#1}{#2} $}}}
\def\0#1#2{\frac{#1}{#2}}

\def\grgl{\:\hbox to -0.2pt{\lower2.5pt\hbox{$\sim$}\hss}{\raise3pt\hbox{$>$}}\:}
\def\klgl{\:\hbox to -0.2pt{\lower2.5pt\hbox{$\sim$}\hss}{\raise3pt\hbox{$<$}}\:}

\begin{document}
\title{Physical renormalization schemes and asymptotic safety in quantum gravity}
\author{Kevin Falls}
\address{\mbox{Institut f\"ur Theoretische Physik, University of Heidelberg, Philosophenweg 12, 62910 Heidelberg, Germany}}

\date{ August 11, 2017 }

\begin{abstract}
The methods of the renormalization group and the $\varepsilon$-expansion are applied to quantum gravity revealing the existence of an asymptotically safe fixed point in spacetime dimensions higher than two. To facilitate this, physical renormalization schemes are exploited where the renormalization group flow equations 
take a form which is independent of the parameterisation of the physical degrees of freedom (i.e. the gauge fixing condition and the choice of field variables).
 Instead the flow equation depends on the anomalous dimensions of reference observables. In the presence of spacetime boundaries  we find that the required balance between the Einstein-Hilbert action and Gibbons-Hawking-York boundary term is preserved by the beta functions. Exploiting the $\varepsilon$-expansion near two dimensions  we consider Einstein gravity coupled to matter. Scheme independence is generically obscured  by the loop-expansion  due to breaking of two-dimensional Weyl invariance. In schemes which preserve two-dimensional Weyl invariance we avoid the loop expansion and find a unique ultra-violet (UV) fixed point. At this fixed point the anomalous dimensions are large and one must resum all loop orders to obtain the critical exponents. Performing the resummation  a set of universal scaling dimensions are found. These scaling dimensions show that only a finite number of matter interactions are relevant. This is a strong indication that quantum gravity is renormalisable.
 \end{abstract}

\maketitle

\section{Introduction}

It remains open problem to identify the theory of quantum gravity which Nature has chosen. 
Due to the dimensionality of Newton's constant $G$
we know that a perturbative quantisation of general relativity does not lead to a predictive theory \cite{Hooft:1974bx,Goroff:1985th}. However, this does not rule out the possibility that quantum version of general relativity may be defined as a local quantum field theory. From the view point of the renormalization group (RG) \cite{Wilson:1973jj,Wilson:1975ly,Polchinski:1983gv,Wetterich:1992yh,Morris:1993qb,Berges:2000ew,Pawlowski:2005xe,Rosten:2010vm}, perturbation theory is just the expansion around the non-interacting low energy fixed point $G=0$; which is simply not the right starting point to formulate the fundamental theory. What we actually require is an interacting UV fixed point $G=G_* \neq 0$  where gravity can be defined as an asymptotically safe theory. At this point  all reaction rates, and other dimensionless observables, remain finite as the UV cut-off is removed \cite{Weinberg:1980gg}.
One may then evade the problem of perturbative non-renormalisability provided there are only a finite number of relevant interactions, i.e. interactions which get stronger as energies decrease. Quantum field theories possessing such a fixed point have been shown to exist for interactions other than gravity \cite{PhysRevLett.55.363,PhysRevLett.36.691}, with the recent example of gauge theories in four spacetime dimensions being of particular interest \cite{Litim:2014uca}.  There is also an increasing amount of evidence in favour of this scenario for quantum gravity in four dimensions, coming from functional renormalization group studies \cite{Reuter:1996cp,Souma:1999at,Lauscher:2001ya,Percacci:2002ie,Litim:2003vp,Codello:2007bd,Machado:2007ea,Niedermaier:2009zz,Benedetti:2009gn,Eichhorn:2009ah,Eichhorn:2010tb,Bonanno:2010bt,Benedetti:2011ct,Donkin:2012ud,Christiansen:2012rx,Henz:2013oxa,Benedetti:2013jk,Falls:2013bv,Dona:2013qba,Codello:2013fpa,Becker:2014qya,Christiansen:2015rva,Meibohm:2015twa,Oda:2015sma,Ohta:2015efa,Demmel:2015oqa,Ohta:2015fcu,Dona:2015tnf,Henz:2016aoh,Falls:2016msz,Christiansen:2016sjn,Eichhorn:2016vvy,Gies:2016con}, and lattice regularisations of quantum gravity \cite{Hamber:1999nu,Ambjorn:2004qm,Laiho:2011ya,Ambjorn:2012jv,Hamber:2015jja,Ambjorn:2016cpa,Laiho:2016nlp}. Additionally, evidence for fixed points in higher than four dimensions has also been found \cite{Fischer:2006at,Fischer:2006fz} using the functional renormalization group.
For reviews on asymptotic safety see \cite{Niedermaier:2006ns,Niedermaier:2006wt,Percacci:2007sz,Litim:2008tt,Reuter:2012id,Reuter:2012xf,Nagy:2012ef}.

A method to study asymptotic safety from within perturbation theory is provided by the $\varepsilon$-expansion around two dimensions  \cite{Gastmans:1977ad,Christensen:1978sc, Kawai:1989yh, Jack:1990ey, Kawai:1992fz, Kawai:1993fq,Kawai:1993mb,Kawai:1995gt,Kawai:1995ju,Kawai:1996vt,Aida:1994dv,Aida:1994zc,Aida:1995ah,Aida:1996zn,Nink:2014yya,Codello:2014wfa,Nink:2015lmq}. In this case one sets the spacetime dimension to $D= 2 + \varepsilon$ where  $\varepsilon$ is a small parameter. The one-loop beta function for Newton's constant then takes the form
\beq \label{betaGintro}
\beta_G = \varepsilon G - b \, G^2\,,
\eeq  
where one expects $b$ to be scheme independent since $G$ is dimensionless in two dimensions. A non-trivial fixed point $G_* = \varepsilon/b + O(\varepsilon^2)$, which can be made arbitrarily small, then presents itself.  Going to higher loops the coefficient $b$ should be replaced by the beta function $\beta_\kappa(D\to2)$ for the inverse Newton's constant $\kappa = 1/G$ obtained in the limit $D \to 2$ by exploiting dimensional regularisation \cite{Bollini:1972ui,tHooft:1973mfk}. One then hopes to resum $\varepsilon$-expansion of the solution to $\beta_G(G_*)=0$ to find a fixed point in integer dimensions $D>2$.\footnote{See e.g. \cite{Morris:2004mg} for the application of this idea to gauge theories in $D>4$ dimensions.}   At this fixed point one would like to know the  scaling dimensions $\theta$ of interactions  to ascertain whether renormalisability can be achieved; this being the case if only a finite number of the exponents $\theta$ have a positive real part.

In exactly two dimensions such critical exponents can be obtained exactly, obeying the Knizhnik-Polyakov-Zamolodchikov (KPZ) scaling relation \cite{Knizhnik:1988ak,Distler:1988jt,David:1988hj}, and the beta function is just given by the conformal anomaly. It was then shown by Kawai,  Kitazawa and Ninomiya \cite{Kawai:1992fz} that the KPZ scaling relation can be reproduced by starting with gravity in $D=2 + \varepsilon$ dimensions and taking a particular limit. However this limit does not correspond to a fixed point in higher dimensions.

Now in the continuum approach to quantum gravity one typically has to choose how to parameterise the physical degrees of freedom. If one calculates observables, i.e. diffeomorphism invariant 
quantities, there should be no dependence on this choice \cite{Kamefuchi:1961sb, Kallosh:1972ap,PhysRevD.13.3247}.  However, in explicit calculations, beta functions appear to depend on the parameterisation via the choice of gauge fixing 
condition and the choice of field variables \cite{Falkenberg:1996bq, Lauscher:2001ya, Litim:2003vp,Percacci:2015wwa,Nink:2014yya,Steinwachs:2013tr,Kamenshchik:2014waa}. This problem leads to apparently scheme dependent value for $b$ (see e.g. \cite{Percacci:2015wwa}) and thus calls into question the physical significance of the fixed point. To make matters worse one also finds a different beta function when the renormalization of boundary terms is considered \cite{Gastmans:1977ad,Christensen:1978sc,Becker:2012js}, leading to an apparently inconsistent theory \cite{Jacobson:2013yqa}. Further to this, going to two-loops appears to produce non-local divergencies spoiling the renormalisability of the theory \cite{Jack:1990ey}.

Our hypothesis is that these problems arise from using renormalization  schemes based on local correlation functions which are 
not themselves observables. Thus, to alleviate this issue one should use a physical renormalization scheme, where we renormalise physical 
observables directly, as was original proposed by Weinberg \cite{Weinberg:1980gg}. The purpose of this paper is to construct such schemes and then use them to resolve the problem of scheme dependence. What we shall see is that generically in $D>2$ dimensions the beta function for Newton's constant can be put into a form which is independent of how we parameterise  the physical degrees of freedom. However, it then depends explicitly on the anomalous dimension of physical observables which reflects the fact that $G$ is dimensionful in dimensions $D>2$.  

We then confront the apparent non-universality of coefficient $b$, obtained in the two-dimensional limit. 
We observe that this problem has its roots in the observation made in \cite{Kawai:1989yh}, namely that the loop expansion close to two dimensions is actually an expansion in $G/\varepsilon$. This has the consequence that $b$ cannot be uniquely determined within a generic scheme and the scaling exponents $\theta$ have order one quantum corrections. The key insight is to observe that the $G/\varepsilon$ expansion is a consequence of schemes breaking two-dimensional Weyl invariance. Using a physical scheme, based on observables that are Weyl invariant in the limit $D\to2$, avoids the expansion in $G/\varepsilon$ and allows for the identification of the fixed point. To calculate the scaling exponents of dimensionful interactions one must then additionally resum the  $G/\varepsilon$ expansion.  After this resummation is performed one has the non-perturbative beta functions which do not suffer from scheme dependence.

We now outline the rest of the paper. We begin by reviewing the formal definition of the functional measure for quantum gravity in section~\ref{sec2}. Several important features are highlighted. In particular we stress that the measure takes diffeomorphism and reparameterisation invariant form which is unique up to an overall normalisation. Furthermore the normalisation is fixed by requiring the absence of non-universal divergencies $\sim \delta(0)$ in the continuum limit \cite{Fradkin:1974df} (see section~\ref{Normalisation_and_local_divergencies}). The two-dimensional limit of the measure is discussed in section~\ref{two_D_limit_measure} and we note that it can be taken in a non-singular fashion provided Newton's coupling also goes to zero as the limit is taken.  In section~\ref{sec3} we discuss the origin of gauge   and parameterisation dependancies when correlation functions are considered.
We note that  these dependencies can be removed by a field renormalization and that certain choices of gauge and/or parameterisation can be understood as giving an implicit renormalization condition for observables. Following from this observation we define physical renormalization schemes in  section~\ref{sec4},   giving the explicit example of schemes based on the volume of spacetime and the volumes of its boundaries. We apply this scheme at one-loop and in general dimension $D>2$ to derive beta functions for Newton's constant and the vacuum energy,  firstly on closed manifolds in section~\ref{sec5}, and then in the presence of  boundaries in section~\ref{sec6}. The beta functions take a form which is independent of the way physical degrees of freedom are parameterised but instead depends on the scaling dimensions of the volumes. In section~\ref{sec7} we consider the beta function for Newton's couplings and matter interactions near two dimensions in a set of schemes based on the renormalization of matter interactions with different classical dimensions. In schemes where the interaction is dimensionless in two dimensions, we argue, in section~\ref{higherloops}, that the loop expansion in $G/\varepsilon$ is avoided and the one-loop beta functions is exact. In section~\ref{2Dquantumgravity} we point out why $G_{\rm IR2D} = -G_*$ is the IR fixed point of two-dimensional quantum gravity which is obtained from higher dimensions where $\varepsilon$ is the IR regulator. We then use the method of Kawai, Kitazawa and Ninomiya \cite{Kawai:1992fz} to resum the expansion in $G/\varepsilon$ using dimensional regularisation and the method of steepest descent. We can then show that the exact beta functions are scheme independent in the two dimensional limit. The explicit form of the non-perturbative scaling dimensions at the UV fixed point in $D>2$ dimensions are also obtained. 
We end with a discussion of our conclusions in section~\ref{conclude}. Several technical steps and results are given in the appendices.

{\it Notation and conventions}: The notation and conventions used in this paper are as follows.

 Greek letters from middle of the alphabet $\mu,\nu...= 0,...,D-1$ are spacetime indices where $D$ is the dimension of spacetime which we take to be $D>2$.  
Lowercase letters from the start of the latin alphabet are DeWitt indices $a,b,c = \{A ,x \}, \{B ,x \}....$ for the fields that parameterise the geometry, and the matter fields when they are present, with the uppercase letters denoting the components (e.g. a symmetric pair of spacetime indices   $A= ( \mu\nu )$ which may be covariant or contravariant) and $x$ denoting the spacetime coordinates  e.g. $\phi^A(x) = g_{\mu\nu}(x)$. Greek letters from the start of the alphabet $\alpha, \beta$ etc. are used for DeWitt indices for the diffeomorphisms e.g. $\xi^\alpha = \epsilon^\mu(x)$.  When we go to a parameterisation where gauge variant and gauge invariant fields are identified $a = \{ \bar{a} , \alpha \}$ where $\bar{a}$ runs over the gauge invariant components and $\alpha$ the gauge variant components.
From the middle  of the latin alphabet $m,n,o = \{M ,x \}, \{N ,x \}....$ are used for super-fields including Fadeev-Popov ghosts e.g. $\varphi^N(x) = \{ g_{\mu\nu}(x), \eta_\mu(x), \bar{\eta}_\nu(x) \}$. 
 When we discuss boundaries $i,j,k,l$ will denote tangential indices and $n$ normal coordinates (no confusion should occur with the DeWitt notation). The covariant derivative with respect to the boundary metric $\ga_{ij}$ is denoted with by $|$ i.e $\ga_{ij|k} = 0$ and $\nabla_{\mu}$ denotes a covariant derivative  with respect to the bulk metric $\nabla_{\rho} g_{\mu\nu} = 0$.

The Einstein sum rule is used throughout and is extended to imply an integral for DeWitt indices  e.g.
\beq
J_{a} \phi^a \equiv \int d^Dx \,  J_A(x) \phi^A(x) \,,
\eeq
and similarly for other indices.
We also use a $\cdot$ to denote ``matrix'' multiplication
\beq
(C \cdot M)_{a}\,^b \equiv  C_{ac} M^{cb} \equiv   \int d^Dy \, C_{AC}(x,y) M^{CB}(y,z)    \,,\,\,\,\,\,\,\,  (C \cdot \phi)_a \equiv  \int d^Dy  C_{AB}(x,y) \phi^B(y)
\eeq
The notation $\det M_{ab}$ denotes the determinant of the matrix $M$ with components $M_{ab}$ and similarly for the super-determinant ${\rm sdet} M_{ab}$ (and similarly for other index sets.).  We use commas and superscripts to denote functional derivatives e.g.
\beq
F^{(1)}_a[\phi] \equiv F_{,a}[\phi]  \equiv \frac{\delta }{ \delta \phi^a} F[\phi]  \equiv  \frac{\delta }{ \delta \phi^A(x)} F[\phi] \,.
\eeq

We work in units where the reduced Planck's constant $\hbar$ and the speed of light $c$ are one.
When we consider renormalization group equations we will typically work in units of the cutoff scale $\Lambda$ and thus it should be understood that all fields and couplings are made dimensionless by the corresponding power of $\Lambda$ such that $\Lambda$ does not appear explicitly in the equations. When we work in units of the cutoff we indicate dependence on $\Lambda$ via the dimensionless RG time
\beq
t  = \log(\Lambda/\Lambda_0)
\eeq
 and derivatives with respect to the cut off will be denoted by $\partial_{t}$. When we work in dimensionful units $\Lambda$ will appear explicitly in expressions and we use $\Lambda \partial_{\Lambda}$ to denote a derivative.

\section{The functional integral in quantum gravity}
\label{sec2}
In this paper we will consider euclidean quantum gravity with spacetime dimension $D> 2$ where we will approach $D \to2$ in a particular limit. 
In order to employ perturbation theory we assume the Wilsonian effective action $S_{\Lambda}$ takes the Einstein-Hilbert form,
\beq \label{EHaction}
S_{\Lambda} \approx S_{\rm EH} =  - \frac{1}{16 \pi G} \int d^Dx \sqrt{g} ( R - 2 \bar{\lambda}  ) + ...
\eeq 
within a semi-classical regime where the cutoff scale $\Lambda$ is sub-Planckian $\Lambda \ll M_{\rm Pl} \equiv G^{\frac{ -1}{D-2}}$. Here $G$ denotes Newton's constant and $\bar{\lambda}$ is the cosmological constant, which is related to the vacuum energy $\lambda$ by $\lambda \equiv \bar{\lambda}/(8 \pi G)$. If the spacetime manifold involves boundaries the action should be supplemented by the required boundary terms denoted by the ellipsis.

Similarly to the action, the functional measure  $d \mathcal{M}(\phi)$ in the sub-Planckian regime should be determined by the canonical quantisation of Einstein's theory. We therefore have the functional integral
\beq \label{Z}
\mathcal{Z} =\int  d \mathcal{M}(\phi) \, e^{-S_{\rm EH} [\phi]} \, , \,\,\,\,\,\,\,\,\,\,\,\,\,\,\,\,\,  
\eeq
where $\phi$ denotes the fields which are being integrated over. 
Here we have not yet introduced the gauge fixing and thus the measure still includes a  formal factor of $V^{-1}_{\rm diff}$ where $V_{\rm diff}$ is the gauge volume which must factor out from the integral over $\phi$.
 In this section we will not concern ourselves with the regularisation of \eq{Z} or the gauge fixing procedure.
Instead the purpose of this section is to find an appropriate formal expression for the measure before regularisation and gauge fixing. 

\subsection{Geometry of geometries}

The fields $\phi$, on which both the action and measure depend, parameterise the (gauge variant) degrees of freedom. Here we assume that they are related to the metric $g_{\mu\nu}$ by an invertible relation
\beq \label{g_to_phi}
g_{\mu\nu} = g_{\mu\nu}(\phi)  \,,\,\,\,\,\,\, \phi^A = \phi^A(g_{\mu\nu})\,,
\eeq
the choice of which cannot affect the physics. Some typical choices for $\phi^A$ are 
\beq
\phi^{A} = g_{\mu\nu}  \,,  \,\,\,\,   \phi^{A} = g^{\mu\nu} \,, \,\,\,\,   \phi^A = \sqrt{g} g^{\mu\nu} \,,
\eeq
which are independent of any background field. With the introduction of a background metric $\bar{g}_{\mu\nu}$ two popular choices for the fields $\phi^A$ are the linear and exponential parameterisations respectively:
\beq \label{linear_expo}
g_{\mu\nu} = \bar{g}_{\mu\nu} + \phi_{\mu\nu}\, , \,\,\,\,\,\,\,\,\,  g_{\mu\nu} = \bar{g}_{\mu\rho} (e^{\phi})^\rho\,_\nu \,,
\eeq
where in the latter case we have the matrix exponential of a field tensor field $\phi$.
While the choice of $\phi$ is unphysical, the geometries which are being integrated over affect the functional integral at least at the non-perturbative level.
This observation motivates the use of the exponential parameterisation \cite{Demmel:2015zfa}, since the positive definiteness of $g_{\mu\nu}$ is ensured even for large values of the field, whereas for the linear parameterisation this is not the case. However, at the perturbative level we do not expect this to be an issue; as we argue below in section~\ref{Integration_limits}.

 In a more general case we can consider variations of the metric
\beq \label{Tcoefficients}
\delta^\n g_{\mu\nu} = \mathcal{T}_{\mu \nu A_1...A_\n}( \phi , \partial_\mu \phi, ... ) \delta \phi^{A_1}... \delta \phi^{A_\n} \,,
\eeq 
where the coefficients $\mathcal{T}$ can depend on the dynamical fields and its derivatives as well as on the background geometry if present. Furthermore, in the most general case  $\mathcal{T}$ are differential operators acting on the variations $\delta \phi^{A}$. Here let us first assume that neither the transformation \eq{g_to_phi} nor the measure $d \mathcal{M}(\phi)$ involve spacetime derivatives. 

Since without any sources present the fields $\phi$ are just integration variables, $\mathcal{Z}$ is invariant under a change 
in the choice of field variables provided we take into account the Jacobian in the measure and re-write $S[\phi]$ in terms of the 
new variables. A useful point of view \cite{Mottola:1995sj} is to consider the fields $\phi^A(x) \equiv \phi^a$ as coordinates on 
the `space of geometries' $\Phi$, to which we associate a metric $C_{ab}[\phi]$ with the line element
\beq \label{line_element}
\delta l^2 = C_{ab}[\phi] \delta \phi^a \delta\phi^b  \,,
\eeq 
which is invariant under a change of coordinates. Thus if we wish to use a different set of field variables $\phi'^a$ which are related to the original variables  $\phi^a=\phi^a[\phi']$ the metric in the coordinate system corresponding to $\phi'^a$ is given by
\beq
 C'_{ab}[\phi']  =  C_{cd}[\phi[\phi']] \frac{ \delta \phi^c}{\delta \phi'^a} \frac{ \delta \phi^d}{\delta \phi'^b}
\eeq
We can write a covariant measure on $\Phi$ as
\beq
d \mathcal{M}[\phi] =  \prod_{a} \frac{d\phi^a}{(2\pi)^{1/2}}   V_{\rm diff}^{-1}[\phi]  \sqrt{ |\det C_{ab}[\phi]| } \,,
\eeq
where  $\sqrt{ |\det C_{ab}[\phi]|}$ provides the volume element  and $V_{\rm diff}[\phi]$ is the volume of gauge orbit corresponding to diffeomorphisms which we take to be a scalar on $\Phi$ such that $V_{\rm diff}'[\phi']= V_{\rm diff}[\phi[\phi']]$. The Jacobian encountered by a change of variables is automatically taken into account by transforming $C_{ab}[\phi]$ since
\bea
 \prod_{a} \frac{d\phi^a}{(2\pi)^{1/2}}   V_{\rm diff}^{-1}[\phi]  \sqrt{ |\det C_{ab}[\phi]| }& =&   \prod_{a} \frac{d\phi'^a}{(2\pi)^{1/2}}  \det \frac{\delta \phi^a}{\delta \phi'^b}  V_{\rm diff}^{-1}[\phi[\phi']]  \sqrt{ |\det C_{ab}[\phi[\phi']]| }\\
 &=&   \prod_{a} \frac{d\phi'^a}{(2\pi)^{1/2}}   V_{\rm diff}^{\prime -1}[\phi']  \sqrt{ |\det C'_{ab}[\phi']| }
\eea
Thus by specifying the form of $C_{ab}$ for a one set of field variables $\phi$ we can then determine the form of the measure in any other set of variables $\phi'$ by determining the the components of  $C'_{ab}$. 

\subsection{Determining the measure}

In principle the metric $C_{ab}$ (or equivalently the measure) of any field can be determined by canonical quantisation \cite{Fradkin:1974df} or by invoking  Becchi-Rouet-Stora-Tyutin (BRST) invariance \cite{Fujikawa:1983im}.
In fact up to an overall normalisation $C_{ab}$ can be determined by demanding that \eq{line_element} is diffeomorphism invariant \cite{Toms:1986sh} which coincides with the BRST invariant form after gauge fixing. On the other hand, Fradkin and Vilkovisky  \cite{Fradkin:1974df}  argue that $C_{ab}$ should be such that the strongest divergencies, which otherwise renormalise the vacuum energy, are removed. They then claim \cite{Fradkin:1976xa} that this is can be achieved by a non-covariant factor of $g^{00}$ entering $C_{ab}$. However, Toms \cite{Toms:1986sh} argues that it is in fact the phase space metric that is non-covariant, leading to a covariant metric on $\Phi$ after integrating out the canonical momentum.

 Following Toms' argument the `correct measure' is that of Fradkin and Vilkovisky but {\it without} the factors of $g^{00}$, which are replaced by  mass scales $\mu^2$ and $\mu^2_{\eps}$. This coincides with the BRST invariant measure of Fujikawa \cite{Fujikawa:1983im} fixing the measure up to an overall normalisation parameterised by $\mu$ and $\mu_{\eps}$. We will employ this form below and then, following Fradkin and Vilkovisky, use the freedom to normalise the measure to remove the strongest divergencies.

To take some simpler examples  \cite{Toms:1986sh} we can consider the quantisation of a single scalar field $s$ (with a canonical action) in curved space where the line element is given by
\beq \label{line_element_s}
\delta l^2  = \mu_s^2  \int d^Dx \sqrt{g} \delta s(x) \delta s(x) \,,
\eeq
which depends on the metric over spacetime via the volume element $\sqrt{g}$ and is invariant under a spacetime diffeomorphism where $g_{\mu\nu}$ transforms as a tensor and $\delta s(x)$ as a scalar.
For a vector field $v^\mu$ we have
\beq \label{line_element_A}
\delta l^2  =  \mu^2_v \int d^Dx \sqrt{g} g_{\mu\nu} \delta v^\mu \delta v^{\nu} \,
\eeq
involving again the metric tensor.
In both cases the form is unique under the assumption that the metric $C$ is ultra-local. 
Here the mass scales $\mu_s$ and $\mu_v$  are needed to ensure that the measure is dimensionless.
We have written \eq{line_element_s} and \eq{line_element_A} in terms of the scalar field $s$ and the vector $v^{\mu}$.
However we can now use a different set of field variables while keeping   $\delta l^2$ invariant. For example instead of using a scalar $s(x)$ we could instead use a density  $\tilde{s} =g^{w/2} s$ of weight $w$   in which case the \eq{line_element_s} is then given by \cite{Toms:1986sh}
\beq
\delta l^2  = \mu_s^2  \int d^Dx \sqrt{g}^{1 - 2w} \delta \tilde{s}(x) \delta \tilde{s}(x) 
\eeq
the choice $w = 1/2$ is then singled out \cite{Fujikawa:1983im} since in this case the metric $C_{ab}$ becomes independent of $g_{\mu\nu}$.  In the case of a vector we can also choose to use densities $\tilde{v}^{\mu} = g^{w/2} v^{\mu}$ or use a one form $v_{\mu}$ instead of a contravariant vector.

Returning to the gravitation degrees of freedom themselves, if we choose $\phi^A(x) = g_{\mu\nu} (x)$ the metric on $\Phi$ is written in the DeWitt form:
\beq \label{deWittmetric1}
C_{ab} \delta \phi^a \delta \phi^b  =  \frac{\mu^2}{32 \pi G} \frac{1}{2}  \int d^Dx \sqrt{g} (g^{\mu\rho}g^{\nu \sigma} + g^{\mu\sigma}g^{\nu\rho} + a   g^{\mu\nu} g^{\rho \sigma}) \delta g_{\mu\nu} \delta g_{\rho\sigma} \,.
\eeq
where $a$ is the DeWitt parameter. Here we will take
\beq \label{a}
a = -1
\eeq
which can be arrived at by several different arguments which we will now briefly recall. Coming from the canonical theory we observe that the projection of this metric with \eq{a} onto a hyper-surface $\Sigma$, with induced metric $\ga_{ij}$, then coincides with the DeWitt metric $\mathcal{G}^{ijkl}  =  \frac{1}{2} \sqrt{\ga} ( \ga^{ik} \ga^{jl} + \ga^{il}\ga^{jk} - \ga^{ij} \ga^{kl})$ appearing in the Hamiltonian:
 \beq
 H = \frac{1}{16 \pi G} \int_{\Sigma} d^{D-1}y \,\left(   \pi^{ij} \mathcal{G}^{-1}_{ijkl}  \, \pi^{kl} - \sqrt{\ga} \, R_{\Sigma} \right)
 \eeq  
 where $\pi^{ij}$ are the canonical momenta and $R_{\Sigma}$ is Ricci scalar on $\Sigma$. Equally, by quantising the theory in a covariant gauge  the measure is determined by the part of the action involving two time derivatives \cite{Fradkin:1974df}. In particular the metric \eq{deWittmetric1} with \eq{a} can be found via
 \beq
g^{00} C_{ab} =  \mu^2 \frac{\delta^2 S_{EH}[\phi]}{\delta \partial_0 \phi^a \delta \partial_0 \phi^b} 
 \eeq
 in Feynman-'t Hooft gauge where the hessian of $S_{EH}$ is a minimal differential operator. Finally, Vilkovisky \cite{Vilkovisky:1984st} also arrives at the same form via arguments based on the connection on $\Phi$ used to define a covariant functional derivative. We therefore take \eq{deWittmetric1} with the DeWitt parameter \eq{a} as defining the measure.

 For the diffeomorphisms we also need a measure in order to define the gauge volume $V_{\rm diff}$. Here again we can choose any parameterisation we like for diffeomorphisms $\xi^\alpha$ since they are only coordinates on the gauge orbit. To be concrete we consider an infinitesimal diffeomorphism 
 \beq
 g_{\mu\nu} \to g_{\mu\nu} + \nabla_{\mu} \eps_\nu +  \nabla_{\nu} \eps_\mu\,,
 \eeq
 then the metric on the space of diffeomorphisms, in these coordinates, is
\beq \label{Gab}
\delta \xi^\alpha \delta\xi^\beta G_{\alpha\beta} =  \frac{\mu^4_{\eps}}{16 \pi G} \int d^Dx \sqrt{g} g^{\mu\nu} \epsilon_{\mu}\epsilon_{\nu}\,,
\eeq
giving the invariant measure for the gauge volume
 \beq
 V_{\rm diff} = \int \prod_{\alpha} \frac{d\xi^{\alpha}}{(2\pi)^{1/2}} \sqrt{ \det G_{\alpha\beta}} \,.
 \eeq
One can then choose a different parameterisation of the gauge orbit transforming $G_{\alpha\beta}$ appropriately.

Some comments are in order.  Here we have assumed that the line element \eq{line_element} and transformations \eq{g_to_phi} do not involve derivatives which then determines the measure by diffeomorphism invariance up to an overall normalisation. However, with the reparameterisation invariant measure we can now make more general, even non-local, field transformations. The important point is that the measure should be ultra-local in a parameterisation $\phi$ which leads to a local action second order in derivatives. Here we consider the unregulated functional integral which is only a formal expression. Once we regulate the theory we will introduce a cutoff scale $\Lambda$ where we will regain the unregulated form of the measure (and/or action)  only in the UV limit $\Lambda \to \infty$.  

\subsection{Normalisation and local divergencies}
\label{Normalisation_and_local_divergencies}
Let's return to the choice of measure and the renormalization of the vacuum energy. It is useful to quote  Fradkin and Vilkovisky \cite{Fradkin:1974df}:

\vspace{2mm}
 ``It is essential for the present discussion that whichever definite, but unique, way of calculating the local measure and the local term in the functional integral is chosen, one will always obtain as a result the cancellation of divergent terms $\propto \delta(0)$ by the local measure."
 \vspace{2mm}
 
As we have defined it, the measure depends on the scales $\mu$ and $\mu_{\eps}$ and thus it is these that we must fix such that the strongest divergencies are removed. Ultimately they will be identified with the cut-off scale $\mu \propto \mu_{\eps} \propto \Lambda$ when the continuum limit is taken. Let us define the relation between the two scales as
\beq
\mu_{\eps}  = \zeta_{\epsilon} \mu
\eeq 
where we treat $\zeta_{\epsilon}$ as a parameter with the ratio between $\mu$ and $\Lambda$ fixed.
Then if we start with $\zeta_{\epsilon} =1$ the effect of shifting the ration $\mu_\eps/\mu \to \zeta_{\epsilon} $ will be to change the normalisation of the functional integral
 \beq \label{FVmeasure}
 \mathcal{Z} \to  e^{- 4 D  \int  d^Dx \delta(0) \log \zeta_{\epsilon} }  \int  d \mathcal{M}(\phi) \,  e^{-S[\phi]} 
 \eeq
 in the continuum limit. This suggests that when the continuum limit is taken we should adjust $\zeta_{\epsilon}$ so that $\mathcal{Z}$ is finite and non-zero e.g. $\mathcal{Z} =1$. 
 If this is not done then there is a factor involving $ \int  d^Dx \delta(0) $ which clearly has no geometrical interpretation, and can be understood as a breaking of general covariance. 
 To elaborate on this point, imagine we want to give meaning to the quantity $\prod_x \zeta^4_{\eps}$ we could do so by writing it as the determinant of some operator 
 \beq
\prod_x \zeta^4_{\eps} = \det \hat{o}
 \eeq
 Written out in components we could then say this operator acts on a scalar like
 \beq
\int d^Dx' \,\hat{o}(x,x') s(x') =  \zeta^4_{\eps}   s(x)
 \eeq
 which leads to \eq{FVmeasure}.
 Now if we consider a diffeomorphism it is evident that $\hat{o}(x,x')$ must transform as a scalar at $x$ and a scalar density at $x'$, but a priori it knows nothing of the dynamical fields. So one must introduce some auxiliary background structure or make some arbitrary choice for the field dependence  $\hat{o} = \hat{o}[\phi]$.

 When the theory is regularised
such divergencies will appear only when the limit $\Lambda \to \infty$ is taken and the form of these divergencies will
depend on $\zeta_{\epsilon}$ which now appears as a parameter of the regularisation scheme. In particular in the regulated theory which preserves diffeomorphism invariance the $ \int  d^Dx \delta(0)$ appear as  divergence proportional to the dimensionful spacetime volume
\beq
\sim \int d^Dx \sqrt{g} \Lambda^D  \,, 
\eeq 
which appears to renormalise the dimensionful vacuum energy by a term proportional to $\Lambda^D$. On the other hand this must follow from some implicit choice of how the operator $\hat{o}$ depends on the dynamical fields in its regulated form. 
Similarly if spacetime boundaries $\Sigma$ are present we will get terms $\sim  \int_{\Sigma} \sqrt{\ga} d^{D-1}y \Lambda^{D-1}$ which renormalise a boundary volume term.   One  should then fix $\zeta_{\epsilon}$ (or more generally the overall normalisation of the measure) in order to remove such divergencies as the continuum limit is taken. This will be possible since such terms are always non-universal.

 We note that in the (causal) dynamical triangulation approaches to gravity such a parameter generally needs to be tuned to uncover phase transitions in four dimensions, either by including a discrete version of \eq{FVmeasure}  in the euclidean version \cite{Laiho:2016nlp}, or by introducing an anisotropy in the regularisation scheme  for causal dynamical triangulations \cite{Ambjorn:2012jv} (which was actually originally advocated by  Fradkin and Vilkovisky \cite{Fradkin:1976xa}). The main point however is not that we must tune a non-universal parameter to obtain a continuum limit, rather we need to tune the parameter if the continuum action is to be of the Einstein-Hilbert form.

\subsection{The two-dimensional limit}
\label{two_D_limit_measure}
Here we have assumed that the dimensionality of spacetime is greater than two.
A key question is whether two-dimensional quantum gravity can be recovered in a particular limit. In two dimensions the Einstein-Hilbert action with a vanishing cosmological constant  $\bar{\la}=0$ is a topological invariant and the classical theory also enjoys Weyl invariance in addition to diffeomorphism invariance.  The Weyl invariance can also be seen in the functional measure since \eq{deWittmetric1} is degenerate in the limit $D\to2$.  In particular if we decompose the metric as $g_{\mu\nu}(x) = e^{2 \sigma(x)} \hat{g}_{\mu\nu}(x)$ where $\hat{g}_{\mu\nu}$ is a uni-modular metric with a fixed determinant  and $\sigma$   parameterises the conformal modes then \eq{deWittmetric1} reads
\beq
C_{ab} \delta \phi^a \delta \phi^b  =  \frac{\mu^2}{32 \pi G} \frac{1}{2}  \int d^Dx \sqrt{g} \left( (\hat{g}^{\mu\rho}\hat{g}^{\nu \sigma} + \hat{g}^{\mu\sigma}\hat{g}^{\nu\rho}) \delta \hat{g}_{\mu\nu} \delta \hat{g}_{\rho\sigma} - 4D (D-2)  \delta \sigma \delta \sigma \right)
\eeq
which reveals that $C_{ab}$ has vanishing eigenvalues in two dimensions. Thus to take the limit $D \to 2$ is problematic. On the other hand if we take also $G \to 0$ while keeping $G/(D-2)$ fixed this limit can be taken since the total measure $\int  d \mathcal{M}(\phi)$ is proportional to factors of $G/(D-2)$. This is the first hint that two-dimensional quantum gravity exists at a fixed point for which $G \propto (D-2)$.

\subsection{Integration limits}
\label{Integration_limits}
Now we return to the question of integration limits; the important point is the following. Imagine we have a standard integral of the form
\beq
\int^{a}_{b}d \phi \sqrt{G^{-1}}  e^{-\frac{1}{G}  \, S(\phi)}\,,
\eeq
where we can think of $G$ as the small parameter in which the integral will be expanded in.
To perform such an integral in perturbation theory one first expands the field $\phi$ about a saddle point $b<\bar{\phi}<a$ and canonically normalises the fluctuations
\beq
\phi \to \bar{\phi} + \sqrt{ G} \delta\phi  \,.
\eeq 
After this the integral is of the form
\beq
\int^{(a- \bar{\phi})\sqrt{1/G}  }_{(b- \bar{\phi})\sqrt{1/G} }d \delta \phi  \, e^{-\frac{1}{G} \, S( \bar{\phi} + \sqrt{ G} \delta\phi)}\,,
\eeq
and we can proceed with the expansion order by order in $G$.
This appears to depend on the limits $a$ and $b$. On the other hand if $G<<1$ we can approximate the integral by
\beq
\int^{(a- \bar{\phi})\sqrt{1/G}  }_{(b- \bar{\phi})\sqrt{1/G} }d \delta \phi \, e^{-\frac{1}{G} \, S(\phi)} \approx \int^{\infty  }_{-\infty }d \delta \phi \, e^{- \frac{1}{G} S( \bar{\phi} + \sqrt{ G} \delta\phi)} \,,
\eeq
where the corrections are exponentially suppressed (i.e. by factors $e^{- {\rm const.} \frac{1}{G}}$) and hence do not contribute to the asymptotic expansion in $G$. 
Evidently the same conclusion is reached at the level of the functional integral since it is just a multiple integral of the same form. Hence the perturbative expansion does not depend on the integration limits for the fields $\phi(x)$.

\section{Gauge and parameterisation dependent beta functions}
\label{sec3}

\subsection{Legendre effective action}

 With the measure in place $\mathcal{Z}$ is manifestly gauge and field parameterisation invariant.
 The problems of gauge and parameterisation dependence arise when we instead consider correlation functions which do not share this property. The first step to obtain correlation functions is to add a gauge fixing action to $S$ along with the corresponding Faddeev-Popov determinant which can be expressed in terms of ghost fields. This step ensures that $\mathcal{Z}$ is unchanged and $V_{\rm diff}$ can be factored out. 
To make this step implicitly let us simply include the ghosts in the set of fields $\varphi^n$ e.g. $ \varphi^n \equiv \varphi^N(x) = \{ g_{\mu\nu},\eta^\mu,\bar{\eta}^\nu\}$ and denote the metric on this enlarged field space by $\C_{nm}$. We then can put the functional integral in the Faddeev-Popov form:
\beq
\mathcal{Z} = \int \prod_{n} \frac{d \varphi^n}{(2\pi)^{1/2}} \sqrt{ |\sdet \, \C_{nm}(\varphi) | } \, e^{-\mathcal{S}[\varphi]} \,,
\eeq 
where $\mathcal{S}[\varphi]= S[\phi] + S_{\rm gf}[\phi] + S_{\rm gh}[\eta,\bar{\eta}, \phi]$ now depends on the gauge fixing condition and both $\mathcal{S}$ and the measure are invariant under BRST transformations.
A typical choice for the gauge fixing action is
\beq \label{gaugefixing}
S_{\rm gf}[\phi] = \frac{1}{32 \pi G \alpha} \int \sqrt{\bar{g}} F^{\mu}(\phi) F_{\mu}(\phi) \,,  \,\,\,\,\,\,   F_{\mu}(\phi) = \bar{\nabla}_{\nu} \phi^{\nu}_{\mu}  - \frac{1}{2} \bar{\nabla}_{\mu} \phi^{\nu}_{\nu}  \,,
\eeq
where the barred quantities depend on the background metric $\bar{g}_{\mu\nu}$.
Since $\mathcal{Z}$ is unchanged it is still independent of the choice of parameterisation and gauge.
The dependence on these unphysical choices enters in the next step in which we couple a source $J_n$ to the fields $\varphi^n$ to obtain
\beq
e^{-\mathcal{W}[J]} =  \int \prod_{n} \frac{d \varphi^n}{(2\pi)^{1/2}} \sqrt{ | \sdet \, \C_{nm}(\varphi) |} \, e^{-\mathcal{S}[\varphi]  + J_n \varphi^n} \,.
\eeq 
 From here one defines the Legendre effective action which is related to $\mathcal{W}[J]$ by a Legendre transformation
\beq \label{LEA}
\Gamma[\bar{\varphi}] =  \mathcal{W}[J] + \bar{\varphi}^n J_n \,,
\eeq
being a functional of the classical fields $\bar{\varphi} = \langle \varphi \rangle_J$ where the subscript denotes that the expectation value is source dependent. 
The functional $\Gamma[\bar{\varphi}]$ is the generating functional for one-particle irreducible correlation functions.
However with $J \neq 0$ the Legendre effective action is neither gauge nor parameterisation independent. By differentiating the effective action we have
\beq
\Gamma_{n}^{(1)}[\bar{\varphi}] = J_n  \,,
\eeq
and consequently it is only when $\Gamma[\bar{\varphi}]$ is evaluated on a solution to the equation of motion that the source is zero.
The off shell action $\Gamma[\bar{\varphi}]$  will therefore depend on both the gauge and the field parameterisation.

\subsection{Origin of gauge and parameterisation dependence}
As mentioned in the introduction the renormalization of Newton's constant at one-loop suffers from unphysical dependencies on the gauge
and parameterisation and furthermore acquires a different form when obtained from bulk or boundary terms in the action.
To  trace the origin of these issues  we now look at how beta functions are typically derived from $\Gamma[\bar{\varphi}] $ at the one-loop level. 
An important point here will be to rebut the claim of \cite{Ohta:2016npm} that: the dependence of beta functions the on parameterisation is physically acceptable and due to the fact that the Jacobian in the path integral measure is not taken into account. Here we will  automatically keep track of the Jacobian by transforming the field space metric $\C_{nm}$ observing that the dependence on the parameterisation is due to the source, rather than the measure.

 Employing the background field method \cite{Abbott:1981ke} the one-loop effective action for gravity can be cast in a gauge invariant (but not independent) form
\beq \label{Gamma}
\Gamma[g_{\mu\nu}] = S[g_{\mu\nu}] + \frac{1}{2} {\rm STr} \log \left( \mathcal{C}^{-1} \cdot \mathcal{S}^{(2)} \right)
\eeq 
where $\Gamma[g_{\mu\nu}] $ is a gauge invariant functional of the metric $g_{\mu\nu}$. In \eq{Gamma} we have combined the contribution from the measure and the Gaussian integrals. 
This expression is only formal since it still needs to be regulated to remove divergencies.
Beta functions can then be found by either demanding that  $\Gamma[g_{\mu\nu}]$ is independent of the UV cutoff $\Lambda$ or by introducing an IR cutoff $k$ on which $\Gamma$ will depend but $S$ is independent of. Note that these are just two different ways of formulating the renormalization group and typically $\Gamma_k$ is related to $S_{\Lambda}$ by a  Legendre transformation \cite{Morris:1993qb,Morris:2015oca} at the exact level.  

Let's now look into the structure of the  operator $\C^{-1} \cdot \mathcal{S}^{(2)}$ involved in the super-trace.
This two point function is made of the product of the inverse field space metric $\C^{-1}$ and the hessian $\S^{(2)}_{nm} \equiv \S_{,nm}$. First let us assume for simplicity that the hessian in a convenient parameterisation is a second order minimal differential operator such that it takes the form:
\beq \label{S2}
\mathcal{S}^{(2)}_{nm} = c_{no} ( -\nabla^2 \delta^o_m -  E^{o}\,_{m}) \equiv  c_{no}  \Delta^{o}\,_m
\eeq
where $c_{nm}$ is structure that appears in front of the Laplacian. The operator $E$  depends on the curvature and the cosmological constant. To understand the dependence of the hessian on the parameterisation we can consider a different set of fields $\tilde{\varphi}$ and find the corresponding hessian. Here we assume that the Jacobian is ultra-local, i.e $\varphi(\tilde{\varphi})$ does not involve derivatives and as such the transformed hessian is again second order in derivatives. One then observers that the hessians are related by:
\bea \label{S2compare}
\tilde{\mathcal{S}}^{(2)}_{nm} &=& \frac{\delta \varphi^o}{\delta \tilde{\varphi}^n}  \S_{op}^{(2)} \frac{\delta \varphi^p}{\delta \tilde{\varphi}^m}  + \frac{\delta \varphi^o}{\delta \tilde{\varphi}^n \delta \tilde{\varphi}^m} \S_{o}^{(1)}  
\eea
where the second term vanishes on the equations of motion $\S_{o}^{(1)} \equiv \S_{,o} = 0$.
On the other hand $ \tilde{\mathcal{S}}^{(2)}_{nm}$  also takes the form \eq{S2} by replacing $c \to \tilde{c}$ and $E \to \tilde{E}$. It follows that the coefficients $\tilde{c}_{nm}$ and $c_{nm}$ are related by 
\beq
\tilde{c}_{nm} = \frac{\delta \varphi^r}{\delta \tilde{\varphi}^n}   c_{rs} \frac{\delta \varphi^s}{\delta \tilde{\varphi}^m}
\eeq
 and thus they transform  as components of metric on field space, just like $\mathcal{C}_{nm}$.
 
 Now the way in which divergencies of \eq{Gamma} are typically regulated is to suppress the modes of $\Delta$ defined in \eq{S2}. However this regulates only the super-trace
 \beq \label{STrDelta}
 \frac{1}{2} \STr \log \left( \Delta \right)
 \eeq
 where we take units $\mu =1$ with $\zeta_{\epsilon} =1$.
As a result there is an unregulated UV divergence  
 \beq \label{delta0}
 \sim \STr \log( \C^{-1} \cdot c) = \delta(0)   \int d^Dx  \, {\rm str} \log( \C^{-1} \cdot c)
 \eeq
 where we have performed the spacetime integral of the super-trace leaving the super-trace   ${\rm str}$ over the indices $A$.
 Usually this divergence is simply neglected, which is justified only if 
  \beq \label{SuperMetric}
{\rm sdet} \, \mathcal{C}_{nm} = {\rm sdet} \,  c_{nm} \,,
 \eeq
 otherwise we will be left with the divergence \eq{delta0}. However, for the BRST invariant functional measure i.e. that based on \eq{deWittmetric1} and \eq{Gab}  one finds that  $\mathcal{C}_{nm} =  c_{nm}$. For example the hessians for the metric and ghosts are given by: 
 \beq
 \frac{\delta^2 \mathcal{S} }{\delta g_{\mu\nu}(x) \delta g_{\rho \sigma}(y)} =    C^{\mu\nu,\rho\sigma}(- \nabla^2 + ...) \delta(x-y) \,,    \,\,\,\,\,\,  \frac{\delta^2 \mathcal{S} }{\delta \eta_{\mu}(x) \delta \bar{\eta}(y)} = G^{\mu\nu}(  - \nabla^2 + ...)  \delta(x-y)
 \eeq
 with the tensor structures those of \eq{deWittmetric1} and \eq{Gab}.
Thus either one adopts the BRST invariant measure which leads to \eq{SuperMetric} or one has additional UV divergencies  unregulated by cutoffs for the modes of the Laplace operator $ -\nabla^2$.
This is inline with Fradkin and Vilkovisky's \cite{Fradkin:1974df} observation  that the correct measure leads to the cancelation of \eq{delta0}.

 
 The main point here is that regulators of the Laplacian do not lead to beta functions dependent on the measure (notwithstanding the field independent normalisation).
To give an example we can add an IR regulator to the one loop expression \eq{Gamma} obtaining:
\beq
\Gamma_k[g_{\mu\nu}] = S[g_{\mu\nu}] + \frac{1}{2} \STr \log \left( \mathcal{C}^{-1} \cdot c \cdot  ( \Delta  +  R_k(-\nabla^2) ) \right)
\eeq
where $ R_k(-\nabla^2)$ is a momentum dependent mass which depends on the IR cutoff scale $k$ such IR modes $p^2 < k^2$ are suppressed. 
 Then taking a $k$ derivative we get the flow equation \cite{Wetterich:1992yh,Morris:1993qb} for the effective average action at one-loop  
 \beq
 k\partial_k \Gamma_k =  \frac{1}{2} \STr \,[  k \partial_k R_k \cdot (\Delta + R_k)^{-1} ]  \,,
 \eeq
 which is studied in \cite{Ohta:2016npm}.
 Since both $\mathcal{C}$ and $c$ fall out of this equation the beta functions will not be depend on them. This shows that the dependence of the beta function on the parameterisation is not due to the functional measure; there is no dependence of $k\partial_k \Gamma_k $ on $\mathcal{C}$. The dependence on the parameterisation instead arises due to the second term in the RHS of \eq{S2compare} which vanishes on shell.

 It is also the off shell corrections to the hessian that introduces the gauge dependence since it is only on shell that the hessian $S^{(2)}$ is guaranteed to be gauge invariant.
 To see this we observe that the action is invariant under
 \beq
 \phi^a \to \phi^a + L^a_\alpha \xi^\alpha 
 \eeq
 for infinitesimal  $\xi^\alpha$ and hence
 \beq
 S_{,a} L^a_\alpha \xi^\alpha  = 0 \,.
 \eeq
Taking a further derivative of the above equation we have
 \beq
  S_{ab}^{(2)} L^a_\alpha \xi^\alpha   +  S_{a}^{(1)} L^a_{\alpha,b} \xi^\alpha = 0
 \eeq 
 which shows that generically only when the equations of motion apply will the quadratic action be  gauge invariant.
 This lack of gauge invariance then leads to the dependence of $\Gamma$ on the gauge fixing condition (see e.g.  \cite{Benedetti:2011ct,Falls:2015qga}).
 We can then conclude that the results of \cite{Ohta:2016npm} still hold even after taking into account the Jacobian and that it is the off shell nature of the calculations
 that is responsible for both gauge and parameterisation dependence.

 \subsection{Gauge and parameterisation dependent beta functions}
 Let's now discuss how the gauge and parameterisation dependence affects the beta functions.
First we can consider the form of the at the UV divergencies which remain relevant in the limit $D\to 2$. They take the form such that the RG equation in terms of {\it dimensionful quantities} form is given by: 
\bea \label{UVdivergencies}
\Lambda \partial_\Lambda S_{\Lambda} &=& \int d^Dx \sqrt{g} \left[  B_0 \Lambda^D +  \Lambda^{D-2} \left( B_1 R  +   \bar{B}_1 \left(R -  \frac{2 D}{D-2} \bar{\lambda} \right) \right) \right]  
\eea
where $\Lambda$ is the UV cut-off scale upon which the couplings depend.
The coefficient of the trace of the Einstein equations $\bar{B}_1$ depends on the gauge and parameterisation whereas the coefficients $B_0$ and $B_1$ are independent of these choices.
More generally, while the one-loop divergencies will be gauge and parameterisation dependent off-shell, going on-shell unphysical dependencies will cancel.
For example such cancellations have been shown explicitly for scalar-tensor theories performed in the Jordon and Einstein frames where the divergencies in the two frames differ off-shell
but agree once the equations of motion are exploited \cite{Steinwachs:2013tr,Kamenshchik:2014waa}.

To obtain the beta functions one goes to dimensionless variables in units of $\Lambda$ such that the metric and couplings  are now dimensionless
(here we take coordinates to be dimensionless $[x^{\mu}] = 0$ from the start and thus $g_{\mu\nu}$ has dimensions of length squared).
To take into account for the modified dimensions we then let $g_{\mu\nu}$ scale as 
  \beq \label{classical_scaling_of_g}
  -  \partial_t  g_{\mu\nu} =  - 2 g_{\mu\nu}\,,
  \eeq 
  which is consistent  with $g_{\mu\nu}$ depending on $t =\log(\Lambda/\Lambda_0)$ as
  \beq \label{classical_scaling_of_g_sol}
  g_{\mu\nu}(t) =  e^{2t} g_{\mu\nu}(0) \,.
  \eeq
  After the transforming to dimensionless quantities we replace the lhs of \eq{UVdivergencies} with
  \beq
 \frac{d}{d t} S[g_{\mu\nu}] =  \partial_{t} S[g_{\mu\nu}] + 2 \int d^Dx g_{\mu\nu}(x) \frac{\delta}{\delta g_{\mu\nu} (x)} S[g_{\mu\nu}]
  \eeq
  where  $\partial_{t}  S[g_{\mu\nu}] \equiv \partial_{t} |_{g_{\mu\nu}} S[g_{\mu\nu}]$ is the partial derivative and thus acts only on the couplings and not on the metric and the second term implements the dilatation step of the RG transformation. In turn in rhs of \eq{UVdivergencies} we simply set $\Lambda =1$ since we are now working in units of the cutoff scale. 
  Then the beta function for the vacuum energy, obtained by keeping track of the terms in  \eq{UVdivergencies}, is given by
\beq \label{beta_lambda_dep}
\beta_{\lambda} = - D \lambda + B_0  -  \frac{2 D}{D-2}   \bar{B}_1 8 \pi G \lambda\,,
\eeq
which depends on the gauge and parameterisation via the last term.
The beta function for Newtons constant depends on whether we use the bulk or boundary term to obtain the running.
In the bulk case we have
\beq \label{beta_G_dep} 
\beta_G = (D-2) G + 16 \pi (B_1+ \bar{B}_1) G^2\,.
\eeq
Let's also note that if we were to obtain the beta function from the boundary action we would obtain a beta function of the form
\beq \label{beta_kappa_boundary} 
\beta_G = (D-2) G + 8 \pi A_1   G^2
\eeq
where for the relative factor between the bulk action and the Gibbons-Hawking-York (GHY) boundary term to be preserved we require that $ A_1 =2 (B_1 + \bar{B}_1)$.
In four dimensions it was shown \cite{Jacobson:2013yqa} that $A_1 =2 B_1$ when employing diffeomorphism invariant boundary conditions. Thus it appears that the relative factor between bulk and boundary is not preserved also due to the term proportional to the equations of motion.

\subsection{Field renormalization and `preferred' parameterisations  }

From the  analysis of this section we can conclude that it is the presence of a source term which leads to gauge and parameterisation dependent beta functions at one-loop. 
Furthermore this may also be responsible for the bulk and boundary terms being renormalised differently. However if we now allow for more general dependence of the field on the cut-off 
 scale we can generate other terms in the renormalization of $S$ proportional to the equations of motion. For example we can allow for an anomalous dimension of the metric by replacing \eq{classical_scaling_of_g} with 
 \beq
 -\partial_{t}   g_{\mu\nu} =  (- 2 + \eta_g) g_{\mu\nu}   \,,
 \eeq 
 which is equivalent to replacing \eq{classical_scaling_of_g_sol} by
 \beq
 g_{\mu\nu}(t) =  e^{(2- \eta_g)t} g_{\mu\nu}(0) \,,
 \eeq
 which leads to an RG equation of the form
 \bea \label{RG_plus_anomolous_term}
\partial_{t}  S_{\Lambda}  + (2 - \eta_g) \int d^Dx g_{\mu\nu}(x) \frac{\delta}{\delta g_{\mu\nu} (x)} S[g_{\mu\nu}] &=& \int d^Dx \sqrt{g} \left[  B_0  +  \left( B_1 R  +   \bar{B}_1 \left(R -  \frac{2 D}{D-2} \bar{\lambda} \right) \right) \right]  
\eea
Thus by choosing the anomalous dimension $\eta_g$ to be non vanishing we can effectively modify the coefficient $\bar{B}_1$, as well as all other coefficients which multiply terms proportional to the equations of motion.    
 One can then use this freedom to satisfy a renormalization condition leading gauge and parameterisation independence beta functions \cite{Kawai:1989yh,Kalmykov:1997uv,Kalmykov:1998cv}.
 Investigating asymptotic safety near two dimensions, using dimensional regularisation, two such renormalization schemes have been proposed \cite{Gastmans:1977ad,Kawai:1989yh}.
 
The first proposal \cite{Gastmans:1977ad}  considered the theory where the cosmological constant was set to zero but the boundary 
terms were retained. There it was argued that the  renormalization of Newton's constant should be determined by divergencies proportional to
\beq
\int d^Dx \sqrt{g} R   +  2  \int_{\Sigma} d^{D-1}y \sqrt{\ga} K 
\eeq
which includes the boundary term,
 rather than the coefficient of $\int d^Dx \sqrt{g} R$ alone, which vanishes on shell. In this case the beta function for Newton's coupling would be identified with \eq{beta_kappa_boundary}. Unfortunately   
 the boundary conditions used in these calculations were not diffeomorphism invariant and therefore the physical significance of the result
 \beq
 \beta_G = (D-2)G - \frac{2}{3} G^2 + O(G^3)\,,
 \eeq
 obtained this way \cite{Gastmans:1977ad,Christensen:1978sc,Becker:2012js} is questionable.  Furthermore redefining the metric will also affect the boundary terms so it is not 
 clear that this method is fully consistent. Nonetheless the general philosophy behind this proposal, which highlights the importance of the boundary terms, should play a role in alleviating the issues surrounding bulk and boundary terms.    
 
 In \cite{Kawai:1989yh} a different renormalization condition was used involving the cosmological constant and other matter couplings where the boundary 
 terms were absent. In this case one can use a redefinition of the metric to remove the terms proportional to the equations of motion in \eq{UVdivergencies} such that a coupling, e.g. the cosmological constant, is not renormalised. 
For the case of the cosmological constant one enforces in dimensionless form
 \beq \label{betalambda0}
\partial_t \lambda = - D \lambda \,,
 \eeq
 where $\lambda$ is dimensionless in units of $\Lambda$.
 When using dimensional regularisation only the logarithmic 
 terms are retained (and hence $B_0=0$) this results in the beta function
 \beq \label{beta_fixed_lambda}
 \beta_G = (D-2)G - \frac{38}{3} G^2 \,.
 \eeq
 It is clear that the requirement \eq{betalambda0} is not unique and one could choose a different condition.
Indeed requiring that different couplings $\g$, other than the cosmological constant, are not renormalised will lead to a different beta function which depends on this choice \cite{Kawai:1989yh}.

More recently several works \cite{Percacci:2015wwa,Falls:2015qga,Gies:2015tca,Benedetti:2015zsw,Ohta:2016npm,Ohta:2016jvw} investigating the the gauge and parameterisation of beta function for Newton's constant  have noted that the dependencies can be minimised by certain choices. In particular one can make use of partial gauge fixings and/or parameterisations such that all additional dependencies are either removed or otherwise satisfy a principle of minimum sensitivity \cite{PhysRevD.23.2916}. To understand why these choices have this effect follows from observing that the beta functions for $G$ and $\lambda$ can be obtained assuming the trace-free Einstein equations hold. As a result the beta functions depend on the gauge and parameterisation  due only to the source for the conformal factor 
\beq
J_{\sigma}(x) \propto R -  \frac{2 D}{D-2} \bar{\lambda}  \,.
\eeq
Here is the field $\sigma(x)$ parameterises conformal fluctuations of the metric such that $g_{\mu\nu} = f(\sigma) \hat{g}_{\mu\nu}$ where the determinant of $ \hat{g}_{\mu\nu}$ is fixed and $ f(\sigma)$ is a function.
The dependence on the source can then be removed either by gauge fixing the conformal factor \cite{Percacci:2015wwa,Benedetti:2015zsw} or picking a parameterisation \cite{Falls:2015qga} where the trace of the Einstein equations does not enter $\mathcal{S}^{(2)}$. In the latter case this can be achieved by choosing a parameterisation where the volume element is linear in the field $\sigma(x)$ 
\beq \label{VolumeParameterisation}
\sqrt{g}(x) - \sqrt{\bar{g}}(x)  = \sigma(x)
\eeq
with $\sqrt{\bar{g}}(x)$ denoting the background volume element. 
 The effect of these choices is that no terms involving the equation of motion appear and hence $\bar{B}_1 = 0$ \footnote{In the case of gauge fixing the conformal factor this can only be done to remove the non-constant modes. As a result the trace of the equation of motion will enter beta functions via the contribution of constant mode $\partial_\mu \sigma_0 = 0$. This does not affect $\bar{B}_1$ in dimensions $D>2$ but will contribute to terms neglected in \eq{UVdivergencies} such as a term $ \sim \bar{B}_2 (R - 4 \bar{\lambda})^2$ in four dimensions. }. Furthermore there is no dependence on the cosmological constant which leads to 
 a real scaling critical exponents for the vacuum energy given simply by its canonical dimension  
 \beq \label{thetalambda}
 \theta_{\lambda} = D \,,
 \eeq   
 obtained in this case by simply differentiating the beta function $\frac{ \partial \beta_\lambda}{ \partial \lambda} = - \theta_{\lambda}$.
 After removing the non-universal divergencies $\sim \Lambda^D$ these gauges then automatically satisfy the renormalization condition that
 the vacuum energy $\lambda$ is not renormalised \eq{betalambda0} in pure gravity. They therefore lead generically to the beta function \eq{beta_fixed_lambda}.

 Although these choices are in some sense preferred it is evident that one should not have to resort to picking specific gauges or parameterisations to get a physically meaningful result. Nonetheless one may wonder whether such gauges or parameterisations implicitly encode more physical information than other choices.
This is the case for the parameterisations \eq{VolumeParameterisation} since they give direct access to the volume of spacetime. To see this note that by integrating the expectation value of $\sigma(x)$ we obtain the volume of spacetime
\beq
\left\langle \int d^Dx \sqrt{g}(x) \right\rangle =  \int d^Dx \sqrt{\bar{g}}(x) +  \int d^Dx \langle \sigma(x) \rangle \,.
\eeq
One can then understand the classical scaling exponent \eq{thetalambda} as expressing the trivial scaling of the spacetime volume
\beq \label{VolumeScaling}
-\partial_{t}  \left\langle \int d^Dx \sqrt{g}(x) \right\rangle = - D \left\langle \int d^Dx \sqrt{g}(x) \right\rangle  \,,
\eeq
and thus for these parameterisation there is an implicit renormalization condition that fixes the  scaling of an observable.
However \eq{VolumeScaling} only applies if $\eta_g=0$ and thus allowing for a non-vanishing anomalous dimension of the metric then leads to a nontrivial scaling dimension for the volume.

 \section{Physical renormalization schemes }
 \label{sec4}
 
 Following from the discussion in the last section we now wish to define { \it physical renormalization schemes} where, instead of any explicit dependence on the  parameterisation of the physical degrees of freedom, the renormalization group equations are written in terms of the scaling dimensions of observables. 
This can be achieved by giving renormalization conditions which relates the renormalization of the fields to the scaling of a set of { \it reference observables}\footnote{Similar ideas have been explored in the context of lattice quantisation of quantum gravity \cite{Cooperman:2014sca}.}. As a result one can maintain both reparameterisation and diffeomorphism invariance (provided of course that they are not broken by regularisation scheme). 
 
To  achieve our aim we work with a regulated functional integral in the absence of sources
\beq
\mathcal{Z} = \int d\mathcal{M}_{\Lambda}[\phi] e^{-S_{\Lambda}[\phi]} \,,
\eeq
where the measure and the action depend on the UV cut off scale $\Lambda$ as indicated by the subscript. This dependence should be such that $\mathcal{Z}$ itself is independent of the scale $\Lambda$, while modes $p^2\gg \Lambda^2$ are suppressed in the functional integral. The RG flow of $S_{\Lambda}$ will then generally encode the coarse graining
of degrees of freedom, renormalization of the fields and a dilatation \cite{Bervillier:2013kda}. Provided we do not break reparameterisation invariance we can then avoid dependence of the choice of parameterisation. Instead, by utilising a physical renormalization scheme, beta functions will depend on the anomalous dimensions of the reference observables. 

By an observable here we mean a function of the fields which is invariant under the symmetries of the theory.
For scalar field theories in flat spacetime the local fields $s(x)$ are observables and as such there renormalization group equations can depend on the anomalous dimension of the fields $\eta_s$ themselves without breaking any symmetry of the theory. If we would consider an $O(N)$ symmetric theory then $s^{\n}$ is not an observable and instead the anomalous dimension must refer to $\varrho = \frac{1}{2} \sum_\n s^{\n} s^{\n}$  which is an observable and hence renormalization group equations can depend on $\eta_{\varrho}$ without breaking the $O(N)$ symmetry.  Since observables are necessarily non-local in quantum gravity we cannot simply identify reference observable with a local function of the fields. Instead here we shall consider observables which are formed by integrating over spacetime or its  boundary $\Sigma$
\beq \label{O_basic_observables}
\mathcal{O} = \int d^Dx \sqrt{g} O(x) \,,  \,\,\,\,\,\,\,\,\,    \mathcal{O} = \int_{\Sigma} d^{D-1}y \sqrt{\gamma} O_{\Sigma}(y)  
\eeq 
for some scalar operators $O(x)$ or $O_{\Sigma}(y)$ which are local functions of the fields. Additionally we can form observables by taking functions of observables of the form \eq{O_basic_observables}.
In physical renormalization schemes the beta functions will depend on the anomalous dimensions which should be self consistently determined at fixed points since there they correspond to universal critical exponents.

\subsection{Volumes as the reference observables}
Let us now give one specific example of such a scheme which we will exploit in the following two sections.
Here we consider the case where we have a compact spacetime manifold with disconnected boundaries $\Sigma_{\m}$.
Then we have classical observables consisting of functions 
\beq \label{VolumeObservables}
\mathcal{O}(\mathcal{V},V_1, V_2, ... )\,,
\eeq
 of the spacetime volume $\mathcal{V} \equiv \int d^Dx \sqrt{g} $ and the volumes of the boundaries $V_\m \equiv \int_{\Sigma_\m} d^{D-1}y \sqrt{\ga}$.
Here the observables \eq{VolumeObservables} will be the reference observables which form the basis of the scheme.
  To this end we consider the renormalization condition
 \beq \label{volume_condition}
-\Lambda \partial_\Lambda \langle \mathcal{O}(\mathcal{V},V_1, V_2, ... ) \rangle = 0\,,
 \eeq
such that the expectation values of the observables \eq{VolumeObservables}  are renormalization group invariants in the absence of any renormalization or dilation of the fields.
This condition can then be understood as a restriction of the RG flow of the Wilsonian effective action which takes the form\footnote{From now on we drop the subscript $\Lambda$ on the action Wilsonian $S$.}
\beq
S =   \lambda(\Lambda) \mathcal{V} +\sum_{\m} \rho_{\m}(\Lambda) V_\m +  \sum_{\n} \g_\n \mathcal{O}_\n\,,
\eeq
with the coupling constants $\lambda$ and $\rho_{\m}$ corresponding to the different volumes respectively and $\mathcal{O}_\n$ denoting the set of all other terms in the action with coupling constants $\g_\n$.
In particular the renormalization condition \eq{volume_condition} can be expressed as the requirement that the RG flow of $S$ is independent of the couplings $\lambda$ and $\rho_\m$,
\beq \label{Flow_condition}
\frac{\partial}{\partial \lambda} \Lambda \partial_\Lambda S = 0= \frac{\partial}{\partial \rho_\m } \Lambda \partial_\Lambda S \,.
\eeq
This follows since then the RG flow of the couplings $\lambda$ and $\rho_\m$ decouples from the flow of all other couplings $\g_\n$ such that the solution to a flow of the type \eq{Flow_condition} involves 
\beq
 \lambda = \lambda(\Lambda_0) +  \int_{0}^{\log(\Lambda/\Lambda_0)} dt \, Y(t; \g_\n(\Lambda_0)) \,,  \,\,\,\,\,\, \rho_\m = \rho_\m(\Lambda_0) +  \int_{0}^{\log(\Lambda/\Lambda_0)} dt \, y_\m(t; \g_\n(\Lambda_0))
\eeq
where $ Y(t; \g_n(\Lambda_0)= \Lambda \partial_{\Lambda} \lambda  $ and $y_\m(t; \g_\n(\Lambda_0)) = \Lambda \partial_{\Lambda} \rho_\m $ are determined from the flow of the essential couplings and $\Lambda_0$ is an arbitrary reference renormalization scale where the boundary conditions for the flow are set.
We then observe that the these couplings are linear in $\lambda(\Lambda_0)$ and $\rho_m(\Lambda_0)$
whereas the couplings $\g_n$ will be independent of  $\lambda(\Lambda_0)$ and $\rho_m(\Lambda_0)$.
Next note that the functional integral can be viewed as a function of the renormalised couplings  
\beq \label{flow}
\mathcal{Z} = \mathcal{Z} (\lambda(\Lambda_0), \rho_\m(\Lambda_0) , \g_\n(\Lambda_0) ) 
\eeq
 which generates the expectation values of observables \eq{VolumeObservables} by taking derivatives with respect to $\lambda(\Lambda_0)$ and $\rho_\m(\Lambda_0)$. For example we obtain the expectation value of the volume via
 \beq
- \frac{\partial}{ \partial \lambda(\Lambda_0)}  \log \mathcal{Z} (\lambda(\Lambda_0), \rho_m(\Lambda_0), \g_\n(\Lambda_0) )  = \langle \mathcal{V} \rangle \,.
 \eeq
Since this is true for any scale $\Lambda$ the condition \eq{volume_condition} follows from the RG invariance of  the functional integral
\beq
\Lambda \partial_{\Lambda} \mathcal{Z} (\lambda(\Lambda_0), \rho_\m(\Lambda_0) , \g_\n(\Lambda_0) )  =0 \,.
\eeq
However taking derivatives with respect to the couplings $ \g_\n(\Lambda_0)$ will not generate the corresponding observable. 
 Thus while the scaling properties of the observables \eq{VolumeObservables} will be trivial the scaling of observables $\mathcal{O}_\n$ will receive quantum corrections.
 
So far we have assumed that the fields do not receive any anomalous scaling and we have not taken the step of rescaling the fields by the cutoff to implement the dilatation step of the RG transformation.
  To regain generality we have to allow for $\phi^a$ to transform under an RG transformation. Without any renormalization of the field the transformation is just a dilatation as in  \eq{classical_scaling_of_g}.
In this case the scaling of the $\langle \mathcal{O}(\mathcal{V},V_1, V_2, ... ) \rangle$ would just give the canonical mass dimension of the observables  fixing the scaling of the observables upon which are renormalization scheme is based. This then limits our search for fixed points unnecessarily \cite{Rosten:2011mf}.
To undo this restriction we can allow for a more general `scaling' of the field which involves quantum corrections to \eq{classical_scaling_of_g} taking the form 
 \beq \label{phiflow}
-  \partial_t \phi^a = d^a[\phi]\,,
 \eeq 
 where $d^a[\phi]$ is some field redefinition
 \beq
 \phi^a \to \phi^a - d^a[\phi] \frac{\delta{\Lambda}}{\Lambda} \,,
 \eeq 
 which can be quite general in principle.
  Here we will assume for the most part that the transformation \eq{phiflow} is a dilatation plus some anomalous scaling given by 
 \beq \label{scaling_of_g_intro}
 - \partial_t g_{\mu\nu} = (- 2 + \eta_g)   g_{\mu\nu}
 \eeq
  where the anomalous dimension $\eta_g = \eta_g(G)$  should vanish at the gaussian fixed point $\eta_g(0)=0$ for $D>2$.
 It then follows that the scaling of the volume is given by
 \beq \label{etaV}
 - \partial_t  \left\langle \int d^Dx \sqrt{g} \right\rangle = d_{\mathcal{V}}  \left\langle  \int d^Dx \sqrt{g} \right\rangle \,,   \,\,\,\,\,\,\,\,\,\,   d_{\mathcal{V}} \equiv - D + \eta_{\mathcal{V}} =  -D + \frac{1}{2} D \eta_g\,.
 \eeq
and similarly for the boundary volumes we have
 \beq \label{etaVb}
 - \partial_t  \left\langle \int d^dy \sqrt{\ga} \right\rangle = d_{V}  \left\langle  \int d^{D-1}y \sqrt{\ga} \right\rangle \,,   \,\,\,\,\,\,\,\,\,\,   d_V \equiv - D +1 + \eta_{V} =  -D +1 + \frac{1}{2} (D -1) \eta_g\,.
 \eeq

If we do not restrict the form of $d^a[\phi]$ a general expression for the scaling of the observables \eq{VolumeObservables} will then be given by 
 \beq \label{ScalingForVolumeObservables}
 -\partial_t  \langle \mathcal{O}(\mathcal{V},V_1, V_2, ... ) \rangle =   \left\langle d^a \frac{\delta}{\delta \phi^a} \mathcal{O}(\mathcal{V},V_1, V_2, ... ) \right\rangle   \,.
 \eeq
  Let us note that this expression for the scaling of the observables has no dependence on the gauge or the parameterisation of the fields.
 This follows since the averages  are being taken without any source term in the functional integral and since $d^a$ transforms as a vector on $\Phi$.
The flow equation should then be of the general form 
 \beq \label{general_flow}
\partial_t  S = d^a \frac{\delta}{\delta \phi^a} S + \mathcal{F}\{S \}   \,,\,\,\,\,\,  {\rm with} \,\,\,\,\,\,   \frac{\partial}{\partial \lambda} \mathcal{F}\{S \} = 0= \frac{\partial}{\partial \rho_\m } \mathcal{F}\{S \} \,,
 \eeq
 where $\mathcal{F}\{S \}$ is the part of the flow equation which represents the coarse graining step of the RG transformation, which depends on the action as indicted by the brackets.
 The first term on the rhs of \eq{general_flow} allows for general field redefinitions \eq{phiflow} which involves a dilatation plus quantum corrections which are of order $G$.
 Thus while the flow equations will now depend on $d^a[\phi]$  its relation to observables is known.
 One then expects that in order to find fixed points where  $\partial_t  S_*[\phi] =0$ we should self consistently determine $d^a_*[\phi]$ leading to a discrete set of physical fixed points as is the case for scalar field theories \cite{Morris:1994ie}.

\subsection{General physical schemes}
In the next two sections we will employ the renormalization scheme based volumes in $D>2$ dimensions. However this is only one physical scheme and one can of course use different schemes for different choices of the reference observables. If we use a set of reference observables  $\mathcal{O}_\m$ with coupling $J_\m$ then we can impose that
 \beq \label{ScalingForObservables}
 -\partial_t  \langle \mathcal{O}_\m \rangle =   \left\langle d^a[\phi] \frac{\delta}{\delta \phi^a} \mathcal{O}_\m \right\rangle   
 \eeq
which leads to a flow equation of the form
\beq \label{flow_for_general_scheme}
\partial_t  S = d^a \frac{\delta}{\delta \phi^a} S + \mathcal{F}\{S \}   \,,\,\,\,\,\,  {\rm with} \,\,\,\,\,\,   \frac{\partial}{\partial J_\m}  \mathcal{F}\{S \} = 0 \,,
\eeq
Close to two dimensions we will exploit a general set of schemes based on observables of different dimensionality. As we shall see this becomes essential to uncover the unique fixed point. Furthermore it is very natural to consider all observables which appear as terms in the action as reference observables. This way one can spot when scheme dependence is broken by an approximation. 

Let us finally note that at the exact level any scheme which is not of the form  \eq{flow_for_general_scheme} but has the form
\beq
\partial_t  S = \tilde{d}^a \frac{\delta}{\delta \phi^a} S + \tilde{\mathcal{F}}\{S \}
\eeq
can still be brought into the form \eq{flow_for_general_scheme}. This will be the case since generically $\tilde{\mathcal{F}}\{S \}$ and $\mathcal{F}\{S \} $ will differ by a term proportional to the equation of motion
\beq
\tilde{\mathcal{F}}\{S \}   =  \mathcal{F}\{S \}    + \Delta d^a \frac{\delta}{\delta \phi^a} S
\eeq
and thus $\tilde{d}^a \to d^a    - \Delta d^a$ restores scheme independence at the exact level. This applies equally to the cases where $\tilde{\mathcal{F}}\{S \}$ is some other physical scheme (i.e independent of some couplings $\tilde{J}_m$) or to generic `unphysical schemes'. Thus at the exact level scheme independence should be preserved \cite{Latorre:2000qc} but, when approximations are made, it may not be possible to see this if information in $\Delta d^a[\phi]$ has been neglected.

\section{One-loop calculation on a closed manifold}
 \label{sec5}

We now consider the case where there are no boundaries present to determine the one-loop running of the vacuum energy and Newton's constant using our renormalization scheme based on the spacetime volume. 
The functional integral takes the form:
  \bea \label{WLambda}
\mathcal{Z} & = &V_{{\rm diff}, \Lambda }^{-1} \int \prod_{a} \frac{d \phi^a}{(2 \pi)^{1/2}} \sqrt{| \det \, C_{ab}^\Lambda(\phi) |} \, \exp\left\{ -\lambda \int d^Dx \sqrt{g}   + \frac{1}{16 \pi G}   \int d^Dx \sqrt{g} R + ... \right\}  \,,
 \eea
 where the ellipsis  denotes terms which enter as loop-corrections not present in the initial action.  The regularisation will be implemented by a modification of the measure $V_{\rm diff}^{-1} C_{ab}(\phi) \to V_{{\rm diff}, \Lambda }^{-1} C_{ab}^\Lambda(\phi)$. The regulated measure is required both to suppress modes $p^2 \gg \Lambda$ and to ensure the renormalization condition \eq{Flow_condition}. Here we do not include the gauge fixing and ghosts in \eq{WLambda} and will instead factor out the gauge volume via a change of variables in the functional integral as we detail in the appendices \ref{A1} and \ref{A2}. A generalisation of \eq{WLambda} in the the presence of spacetime boundaries will be given in section~\ref{sec6}.

\subsection{Perturbative expansion and regularisation}

 To compute $\mathcal{Z}$ to leading order in $G$  we make the split
 \beq
 \phi^a = \bar{\phi}^a + \delta\phi^a \,,
 \eeq
 expanding the integrand of \eq{WLambda} around the saddle point $\bar{\phi} = \bar{\phi}(\lambda,G)$ which depends on the couplings. 
 It follows that the saddle point geometry must be an Einstein space where the Ricci curvature 
 \beq \label{curvature}
R_{\mu\nu}(\bar{\phi}) = g_{\mu\nu}(\bar{\phi}) \frac{16 \pi G }{D-2} \lambda \,,
\eeq
depends explicitly on the couplings. Since $\lambda$ is related to the curvature we can then avoid counter terms in the RG flow that depend on $\lambda$ and hence satisfy \eq{Flow_condition} by renormalising curvature dependent terms instead.
This allows us to implement \eq{Flow_condition} at each order in perturbation theory if we do not include any anomalous dimension for the metric.  

To obtain the one-loop quantum corrections we have to compute the Gaussian integral over the gauge invariant modes by first extracting the gauge orbit from the integral over the gauge variant fields.
 This can be done by fixing the gauge and is most easily achieved by adopting the Feynman-'t Hooft gauge \eq{gaugefixing} where $\alpha =1$. However it is possible to factor out the gauge orbit without fixing the gauge \cite{Mazur:1989by,Bern:1990bh, Mottola:1995sj,Vassilevich:1995bg} but instead using the freedom to pick coordinates $\phi^a$ which split the field into physical and gauge degrees of freedom. Gauge independence is then just reflected in the fact that appropriate coordinate systems, corresponding to different gauges, are just related by transformations with a trivial Jacobian.   This procedure is outlined in Appendix~\ref{A1} and the resulting determinants, along with the Gaussian integrals, are evaluated explicitly in Appendix~\ref{A2}. The same result can be obtained from the standard Faddeev-Popov gauge fixing procedure \cite{Falls:2015qga} apart from a complete treatment of zero modes which we treat here in detail (see  Appendix~\ref{A3} and \cite{Volkov:2000ih}). 
 The final result is manifestly gauge independent and is invariant under field reparameterisations:
  \beq \label{unregulated}
 - \log \mathcal{Z} = S_{\Lambda} [\bar{\phi}] + \frac{1}{2} \Tr_{2} \log( \Delta_2/\mu^2)  - \Tr_{1}' \log (\Delta_1/\mu_{\eps}^2)   + \log \Omega(\mu_{\eps})    \,.
 \eeq
 where here all quantities evaluated at the saddle point \eq{curvature}.
The differential operators $\Delta_1$ and $\Delta_2$   act on vectors and symmetric tensors respectively and are given by
\beq \label{Deltas}
\Delta_1 \epsilon_\mu =  \left( - \nabla^2 - \frac{R}{D} \right)  \epsilon_\mu  \,,\,\,\,\,\,\,\,  \Delta_2 h_{\mu\nu} = -\nabla^2 h_{\mu\nu}  - 2 R_\mu\,^\rho\,_\nu\,^\sigma h_{\rho \sigma}  \,,.
\eeq
 The prime indicates that the zero modes should be removed from the vector trace. These correspond to Killing vectors  i.e. the subgroup of  diffeomorphisms $\mathcal{H}$ which are isometries of the saddle point geometry $\bar{\phi}^A$. The invariant volume $\Omega(\mu)$ on  $\mathcal{H}$ (given explicitly by \eq{Omega} in Appendix~\ref{A3}) then appears in the last term of \eq{unregulated} to ensure these modes are removed from the functional integral.

Since \eq{unregulated} is divergent we need to regulate the traces.
Our regularisation procedure is implemented at the level of the measure via a modification of the field space metrics $C_{ab}$ and $G_{\alpha \beta}$ which implements a proper-time regularisation. Working in dimensionful units the explicit form of  the regulated measure can be expressed in terms of the metric
  \beq \label{CLambda}
C_{ab}^{\Lambda} \delta \phi^a \delta \phi^b =    \frac{1}{32 \pi G} \int d^Dx \sqrt{g} \frac{1}{2} \left(  g^{\mu\alpha}g^{\nu\beta} + g^{\mu\beta}g^{\nu\alpha} -  g^{\mu\nu} g^{\alpha \beta} \right) \delta g_{\mu\nu}    \,\Delta_2 e^{\gamma(\Delta_2/\Lambda^2)}  \,\delta g_{\alpha\beta} \,,
  \eeq 
while the metric on the space of diffeomorphisms \eq{Gab} is replaced by
  \beq \label{GLambda}
G_{\alpha \beta} \xi^\alpha \xi^\beta =  \frac{1}{16 \pi G} \int d^Dx \sqrt{g} g^{\mu\nu} \delta \epsilon^\mu  \, \Delta_1^2 e^{2\gamma(\Delta_1/(\Lambda^2 \zeta_{\epsilon}^2)}  \, \delta \epsilon^\nu  \,,
  \eeq
     where:
  \beq
  \gamma(z) \equiv \int_{1}^\infty  \frac{ds}{s} e^{-s z} \,,
  \eeq
  is the  incomplete gamma function. 
  Here the measures depends on the dynamical fields $\phi$ rather than the saddle point geometry which is necessary for the renormalization condition \eq{Flow_condition} for $S[\phi]$.

     This regularisation ensures that $\mathcal{Z}$ is UV regulated at one-loop order, 
  in particular it has the effect to replace \eq{unregulated} by the regulated expression
 \beq \label{W1reg}
  -\log \mathcal{Z} = S_{\Lambda}[\bar{\phi}]  -\left( \frac{1}{2} \Tr_{2} \,  \gamma(\Delta_2/\Lambda^2)  - \Tr_{1}' \,  \gamma(\Delta_1/(\Lambda^2 \zeta_\eps^2))  \right)  + \log \Omega( \zeta_{\epsilon} \Lambda   e^{-\gamma_E/2}) \,.
 \eeq
 where all field dependent quantities are evaluated on the source dependent saddle point and here $\gamma_E$ is Euler's constant.
  We then observe that for low momentum modes 
\beq
- \gamma(\Delta_2/\Lambda^2 \to 0 )  = \log( e^{\gamma_{E}} \Delta_2/\Lambda^2 )\,, \,\,\,\,\,   - \gamma(\Delta_1/(\Lambda^2 \zeta_\eps^2))  = \log( e^{\gamma_{E}} \Delta_1/(\Lambda^2 \zeta_{\eps}^2) ) \,,
\eeq
which is of the form \eq{unregulated} with 
\beq
\mu^2 = \Lambda^2   e^{-\gamma_E} \,,
\eeq
and $\mu_{\eps} = \zeta_{\eps} \mu$. 
For high momentum modes we have
\beq
- \gamma(\Delta_2/\Lambda^2 \to \infty ) =  \frac{\Lambda^2}{\Delta_2} e^{-\Delta_2/\Lambda^2}\,,  \,\,\,\,\,\,\,   - \gamma(\Delta_1/(\Lambda^2 \zeta_\eps^2) \to \infty ) =  \frac{\Lambda^2 \zeta_\eps}{\Delta_1} e^{-\Delta_2/(\Lambda^2\zeta_\eps^2)} \,,
\eeq
which vanishes exponentially quickly such that $\mathcal{Z}$ is finite.
As such the modified measure regulates the one-loop divergencies while introducing the cut-off scale $\Lambda$. 
Sending $\Lambda \to \infty$ the measure returns to the unregulated form as required. 
 

\subsection{One-loop flow equation}
  We now want to calculate the RG flow of $S[\phi]$ where we will incorporate a renormalization of the fields and a dilation.
 Let us denote the overall volume element in modified functional measure  as 
 \beq
 \mathcal{M} =  V_{{\rm diff}, \Lambda }^{-1} \sqrt{| \det \, C_{ab}^\Lambda |} \,,
 \eeq
then it is straight forward to show that before renormalization of the fields we have
\beq
\Lambda \partial_\Lambda  \mathcal{M}   =   \left( \Tr_2 [e^{- \Delta_2/\Lambda^2 }  ] - 2 \Tr_1 [e^{- \Delta_1/(\Lambda \zeta_{\epsilon})^2 }  ] \right) \mathcal{M}  +   O(G)
\eeq
with the right hand side given by the trace of the heat kernels. Here we made use of the scaling property \eq{LambdaOmega} of $\Omega$, which implies  $\Lambda \partial_\Lambda \log \Omega(\Lambda \zeta_{\epsilon}   e^{-\gamma_E/2}) = 2 N_{KV}$,  to absorb this contribution into the vector trace by dropping the prime. As a result the flow equation is unaffected by the number of Killing vectors.

Now when we go to scaled and renormalised fields we absorb $\Lambda$ into the fields and use dimensionless couplings such that $G$ and $\lambda$ now denote couplings in units of $\Lambda$.
 Then we have that the measure for the scaled and renormalised fields scales according to 
\beq \label{Measureflow}
 \frac{d}{d t}   \mathcal{M}  \equiv \partial_t   \mathcal{M} - d^a \frac{\delta }{\delta \phi^a}  \mathcal{M}      =  \left( \Tr_2 [e^{- \Delta_2 }  ] - 2 \Tr_1 [e^{- \Delta_1/ \zeta_{\epsilon}^2 }  ]   +  \frac{\delta d^a }{\delta \phi^a}  \right) \mathcal{M} +   O(G)
\eeq
where the last term accounts for the the Jacobian picked up when transforming to the scaled and renormalised fields and we again drop terms of order $G$.
Then we note that exact RG equations follows from \cite{Latorre:2000qc}:
\beq \label{exactRG}
\partial_t   (\mathcal{M}e^{-S})  = \frac{\delta}{\delta \phi^a} ( \Psi^a \mathcal{M}e^{-S}) 
\eeq
for some choice of $\Psi^a$ giving different schemes.
The invariance of $\mathcal{Z}$ follows since the integral of \eq{exactRG} is zero. Note that this implicitly 
sets the boundary of integration for the functional integral since we must have that $\Psi^a \mathcal{M}e^{-S}$ vanishes on the boundary.
Here set $\Psi^a = d^a$ to obtain the one-loop flow equation
\beq \label{oneloopflow}
\partial_t  S
 =    d^a \frac{\delta}{\delta \phi^a} S   +  \Tr_2 [e^{- \Delta_2 }]   - 2 \Tr_1 [e^{- \Delta_1/\zeta_{\eps}^2 }  ]   \,.
   \eeq
which is of the form \eq{general_flow} with 
\beq \label{Fpropertime}
\mathcal{F} =  \Tr_2 [e^{- \Delta_2 }]   - 2 \Tr_1 [e^{- \Delta_1/\zeta_{\eps}^2 }  ]  \,,
\eeq
and \eq{ScalingForVolumeObservables} follows by integrating by parts.
Note that in principle any term proportional to the equation of motion can be removed from \eq{oneloopflow}  by a specific choice of  $d^a$ however the repercussion of such a choice is to induce a non-trivial scaling \eq{ScalingForVolumeObservables} for observables which depend on the volumes.
For our choice of regularisation the flow equation has the form of a proper-time flow but with the additional term that accounts for the renormalization of the fields.
Proper-time flows have been studied previously in the context of asymptotic safety \cite{Bonanno:2004sy,Bonanno:2005bj}. Here we stress that these flow equations only regulate the one-loop divergencies. Later we will exploit dimensional regularisation to go beyond one-loop. 

The point to recognise is that in the one-loop approximation we can choose any regulator which regulates the gaussian integral which is performed at the saddle point. This decides that the differential operators which appear in \eq{CLambda} and \eq{GLambda} are given by \eq{Deltas} when evaluated for the saddle point geometry $\phi =\bar{\phi}$.
  The additional renormalization condition \eq{Flow_condition} then decides that for general $\phi$ they are independent of the vacuum energy.
 It is then ensured that the dependence of  $ \Lambda \partial_{\Lambda} S$ on $\lambda$ comes from the first term in \eq{oneloopflow} and hence $d^a$ is related to the anomalous scaling of the volume via \eq{ScalingForVolumeObservables}. We could add to the differential operators \eq{Deltas} terms involving the trace-free Ricci tensor $S_{\mu\nu} \equiv R_{\mu\nu} - \frac{1}{D} g_{\mu\nu} R$ since for the saddle point $S_{\mu\nu}(\bar{\phi}) = 0$. These won't modify the renormalization of Newton's coupling however.

 \subsection{One-loop beta functions}
 
Expanding the heat kernel for the operators \eq{Deltas} in the early-time expansion we obtain 
 \bea
  \Tr_2 [e^{- \Delta_2}  ] - 2 \Tr_1 [e^{- \Delta_1 }  ]  &=&     \frac{ \frac{1}{2}D(D+1)  - 2 D \zeta_{\epsilon}^D }{(4 \pi )^{\frac{D}{2}}} \int d^Dx \sqrt{g}  \nonumber  \\
   &&+  \frac{1}{6} \frac{\left(\frac{1}{2}D(D+1)  - 6 -  ( 2D + 12) \zeta_{\epsilon}^{D-2} \right)}{(4 \pi )^{\frac{D}{2}}}  \int d^Dx  \sqrt{g} R  + ... \,.
 \eea
 Then acting the dilatation operator on the action and allowing for an anomalous scaling of the metric \eq{scaling_of_g_intro} we have
\beq
 d^a \frac{\delta}{\delta \phi^a} S = - (- 2 + \eta_g) \frac{1}{2}   \left(  D \lambda \int d^Dx \sqrt{g} - (D-2) \frac{1}{16 \pi G}  \int d^Dx \sqrt{g} R \right)
\eeq
where $\eta_g$ is the anomalous dimension of the metric.
The flow equation \eq{oneloopflow} then leads to the beta functions for the dimensionless couplings
\bea\label{betas}
\beta_G &=& (D-2)\left( 1 - \frac{\eta_{\mathcal{V}}}{D} \right) G - \frac{2 }{3} \frac{\left(\frac{1}{2}D(D+1)  - 6 -  ( 2D + 12) \zeta_{\epsilon}^{D-2} \right)}{(4 \pi)^{\frac{D-2}{2}}} G^2 \,, \\
\label{betas_lambda}
  \beta_\lambda &=& \left( - D  + \eta_{\mathcal{V}} \right)\lambda +  \left( \frac{1}{2}D(D+1)  - 2 D \zeta_{\epsilon}^D \right) \frac{1}{(4 \pi )^{\frac{D}{2}}} \,,
\eea 
which are completely independent on the gauge or parameterisation and instead are written in terms of the anomalous scaling dimension of the volume $\eta_{\mathcal{V}}$ given by \eq{etaV}. Note that since $\eta_{\mathcal{V}}$ must vanish at the Gaussian fixed point it must be order $G$ 
\beq
\eta_{\mathcal{V}} = G \eta_{\mathcal{V},1} + ... \,,.
\eeq 
where $\eta_{\mathcal{V},1} $ is a constant.
For $\zeta_{\epsilon} =1$ and $\eta_{\mathcal{V}}=0$ the beta functions \eq{betas} agree with \cite{Falls:2015qga}. Here we see that the beta functions take a more general form in terms of the anomalous dimension and the measure parameter  $\zeta_{\epsilon}$. Note that in the limit $D \to 2$ the beta function for Newton's constant becomes independent of $\zeta_{\epsilon}$.
Ultimately the value of $\zeta_{\epsilon}$ should be fixed in the continuum limit. If we only consider the one-loop beta functions its value should be such that the constant term in $\beta_{\lambda}$ vanishes which leads to the value
\beq
\zeta_{\epsilon}^{\rm crit} = \frac{1}{4^{\frac{1}{D}}} (1 + D)^{\frac{1}{D}}
\eeq
which is of order one for all $2 < D < \infty$ and is given by $\zeta_{\epsilon}^*  =\frac{\sqrt{3}}{2}$ in the limit $D \to 2$.  
For this choice of  $\zeta_{\epsilon}$ there exists a fixed point for which $\lambda = 0$.

\subsection{Discussion}
The beta functions \eq{betas} and \eq{betas_lambda} are independent of any gauge fixing parameters and the parameterisation of the quantum fields by virtue of our approach based on observables.
However they do depend on the anomalous dimension of the reference observable which we have chosen to be the spacetime volume. 
The situation here is similar to that of scalar field theories where renormalization group equations will depend on the anomalous dimension of the fields.
While at fixed points we expect that anomalous dimensions should be scheme independent, provided they do not correspond  to the scaling of a redundant operator \cite{Wegner1974}\footnote{Recall that a redundant operator corresponds to any coupling that is an eigen-perturbation of a fixed point which can be removed by a field redefinition. For a discussion on redundant operators in the context of quantum gravity see \cite{Dietz:2013sba}. }, away from fixed points the anomalous dimensions are scheme dependent and as such they appear in the one-loop beta functions as undetermined parameters. In section~\ref{sec7} we shall see how the anomalous dimensions can be determined at the the UV fixed point in $D =2 + \varepsilon$ dimensions.  

Given the dependence of the one-loop beta functions on $\eta_{\mathcal{V}}$ one may ask whether there is a quantity which is independent of $\eta_{\mathcal{V}}$.
Since the volume is dimensionful one can in principle only measure its ratio with an other dimensionful scale to form a dimensionless number. This suggests that we consider the spacetime volume measured in Planck units
\beq
\tilde{\V} = G^{- \frac{D}{D-2}} \V\,,
\eeq
then if we consider its scaling we obtain
\bea
-  \partial_{t}  \langle \tilde{\V}  \rangle &=& \frac{D}{D-2} \frac{\beta_G}{G} \langle \tilde{\V}  \rangle + (-D + \eta_V)  \langle \tilde{\V}  \rangle \nonumber \\
&=&   - \frac{2 }{3} \frac{D}{D-2}  \frac{\left(\frac{1}{2}D(D+1)  - 6 -  ( 2D + 12) \zeta_{\epsilon}^{D-2} \right)}{(4 \pi)^{\frac{D-2}{2}}} G \langle \tilde{\V}  \rangle\,,
\eea
which is indeed independent of the anomalous dimension of the volume.

\section{Amplitudes and the renormalization of boundary terms}
 \label{sec6}
 
In the preceding section we assumed that the spacetime manifold had no boundary.
We now wish to consider the case where we have a boundary which allows for us to compute amplitudes
\beq \label{Amplitude}
\langle \phi_1 | \phi_2 \rangle = \mathcal{Z}[\phi_1, \phi_2] \,,
\eeq
where $\phi_1$ and $\phi_2$ denote boundary data which constrains the fields on the two boundaries $\Sigma_1$ and $\Sigma_2$.
Provided these boundary conditions are diffeomorphism invariant they correspond to different quantum states and $ \mathcal{Z}[\phi_1, \phi_2]$ constitutes a physical observable i.e. an amplitude in the physical Hilbert space.
Subject to these boundary conditions the action must be supplemented with boundary terms \cite{York:1972sj,Gibbons:1976ue} such that the action has
a meaningful variational principle and amplitudes have the required composition properties \cite{Hawking:1980gf}. This typically leads to a requirement that
the bulk and boundary terms be interrelated. 

\subsection{Action and boundary conditions}

Quantum gravity on manifolds with boundaries faces a problem \cite{Avramidi:1997sh} due to the generic lack of diffeomorphism invariant boundary conditions which lead to a well defined heat kernel for differential operators, such as $\Delta_1$ and $\Delta_2$. However, such boundary conditions \cite{Moss:1996ip,Moss:2013vh} do exist for geometries where the extrinsic curvature $K_{ij}$ on the boundary $\Sigma$ takes the form 
 \beq \label{Kij}
 K_{ij} =\frac{1}{D-1} K \, \ga_{ij} \,,\,\,\,\,\,\,\,\,\,\, \partial_i K = 0 \,,
 \eeq
 where $i,j$ etc. denote tangential coordinates, $\ga_{ij}$ is the induced metric and $K= \ga^{ij} K_{ij}$.
 Explicitly these boundary conditions are given by \cite{Jacobson:2013yqa,Moss:1996ip,Moss:2013vh}:
 \bea  \label{BRST_bcs}
 h_{in} &=& 0 = \epsilon_n \\
  \dot{\epsilon}_{i} - K^j_i \epsilon_j &=& 0 \\
\dot{h}_{nn} + K h_{nn} -  2 K^{ij} h_{ij}& = &0  \\
\dot{h}_{ij} - K_{ij} h_{nn} &=& 0 
\eea
where the dot is a normal derivative and $n$ denotes the normal components components of tensors $h_{\mu\nu} = \delta \phi^A$ and vectors $\epsilon_{\mu}$ on which $\Delta_2$ and $\Delta_1$ act. One can explicitly check that these boundary conditions are gauge invariant under the transformation
\beq
h_{\mu\nu} \to h_{\mu\nu} + \nabla_\mu \epsilon_\nu + \nabla_\nu \epsilon_\mu
\eeq
provided $K_{ij}$ takes the form \eq{Kij}. Some important results concerning the application of these boundary conditions as well our conventions  
are given in Appendix~\ref{boundaryApp}.

When we make loop expansion of the amplitude \eq{Amplitude} the boundary conditions \eq{BRST_bcs} are to be imposed on fluctuation fields $\delta \phi^A = h_{\mu\nu}$ where the saddle point $\bar{\phi}^A$ is a geometry with extrinsic curvature \eq{Kij}.
We therefore seek an action which has an extremum for such a geometry while giving rise to the linearised Einstein equations for $\delta \phi^A$. To this end we consider the action
\bea \label{Sb}
S =   \lambda \int d^Dx \sqrt{g}     
 - \frac{1}{16 \pi G}  \left( \int d^Dx \sqrt{g} R   +  2  \int_{\Sigma} d^{D-1}y \sqrt{\ga} K \right) +   \rho \int_{\Sigma} d^{D-1}y \sqrt{\ga}   
\eea
where we have included the Gibbons-Hawking-York (GHY) boundary term as well as a boundary term corresponding to the volume of the boundary (here we include a single boundary for simplicity). Expanding the action via $g_{\mu\nu} \to g_{\mu\nu} + h_{\mu\nu}$ we obtain:
\beq
S = S[g_{\mu\nu}] +  \int d^Dx \sqrt{g}\,  \mathcal{E}^{\mu\nu} h_{\mu\nu}    + \frac{1
}{2}  \int d^{D-1}y  \sqrt{\ga}  \left( \rho  \ga^{ij}  + 2 \frac{1}{16 \pi G} (K^{ij}  -  K \ga^{ij})  \right) h_{ij}  \nonumber + ...
\eeq
where $\mathcal{E}^{\mu\nu} =   \frac{1}{16 \pi G} R^{\mu\nu}  - \frac{1}{16 \pi G} \frac{1}{2} R g^{\mu\nu}  + \frac{1}{2}  \lambda g^{\mu\nu}$ denotes the Einstein field equations. One then observers that the boundary terms vanishes provided the extrinsic curvature evaluated on the saddle point $\bar{\phi}^A$  is given by \eq{Kij} with
\beq
K= \frac{D-1}{D-2}  8 \pi G \rho \,,
\eeq
whereas the bulk term vanishes for a solution to the Einstein field equations.

Computing the action to quadratic order in the fluctuation $h_{\mu\nu}$ around this background and applying the boundary conditions \eq{BRST_bcs} (along with the identities given in  Appendix~\ref{boundaryApp}) one finds that all boundary terms vanish and we obtain the linearised Einstein field equations for $h_{\mu\nu}$. The hessian is gauge invariant having the same form as the one obtained without a boundary, in particular we recover \eq{delta2S}, \eq{hTThess} and \eq{shess}. 
If instead of \eq{Sb} a different relative coefficient for the GHY term is chosen the hessian involves involves boundary terms and hence we cannot use such an action to derive the linearised Einstein equations around an on shell background. 

\subsection{Functional integral}

It follows that we may generalise our one-loop calculation including the boundary terms with the functional integral now given by:
\bea \label{WLambda_b}
\mathcal{Z}[\phi_1, \phi_2]   & = &V_{{\rm diff}, \Lambda }^{-1} \int \prod_{a} \frac{d \phi^a}{(2 \pi)^{1/2}} \sqrt{ \det \, C_{ab}^\Lambda(\phi) } \, \exp\left\{ - \lambda \int d^Dx \sqrt{g}  -  \rho_1  \int_{\Sigma_1} d^Dy \sqrt{\ga}     -  \rho_2 \int_{\Sigma_2} d^Dy \sqrt{\ga}         \right.  \nonumber \\  
&& \left.    + \frac{1}{16 \pi G}   \left( \int d^Dx \sqrt{g} R   +  2  \int_{\Sigma_1} d^{D-1}y \sqrt{\ga} K  +  2  \int_{\Sigma_2} d^{D-1}y \sqrt{\ga} K  \right)   +...\right\} 
 \eea
Where we include two separate boundaries to give the interpretation of $W = \log Z[\phi_1, \phi_2]$ as a an amplitude with the total boundary being the disjoint union $\Sigma = \Sigma_1 \cup \Sigma_2$.
To compute $\mathcal{Z}$ at one-loop we proceed as before but now the saddle point geometry has extrinsic curvature
\eq{Kij} with 
\beq \label{KJ}
K_{\Sigma_{1,2}}(\bar{\phi}) =  \frac{D-1}{D-2}  8 \pi G \rho_{1,2}  \,,
\eeq
dependent on the couplings. Our requirement that the counter terms do not involve the couplings $\rho_1$ and $\rho_2$ can be satisfied by renormalising terms which depend on  $K(\phi)$ in a similar manner to how the dependence on $\lambda$ is evaded.
It follows from \eq{KJ} that the boundary data $\phi_1$ and $\phi_2$ corresponds to defining
\beq
\phi_{1,2}^A = \bar{\phi}_{1,2}^A + \delta \phi_{1,2}^A
\eeq
and requiring that the backgrounds $\bar{\phi}_{1,2}$ have extrinsic curvature \eq{Kij} with \eq{KJ} fixing the constant background $K$ on each boundary and the boundary conditions of the fluctuations given by \eq{BRST_bcs}. Since by varying $\rho_1$ and $\rho_2$ we can set different values $K_{\Sigma_{1}}$ and $K_{\Sigma_{2}}$ we have access to a two parameter family of amplitudes. Importantly the steps needed to calculate  $\mathcal{Z}$ to one-loop,  detailed in Appendix~\ref{A2}, can be carried out with the boundaries present.
The non-local operators appearing in the measure are then defined with the boundary conditions \eq{BRST_bcs}.

\subsection{One-loop RG flow with boundary terms} 

 The flow equation then takes the form \eq{oneloopflow} but where now the heat kernels are subject to the boundary conditions \eq{BRST_bcs} leading to an RG flow for the boundary terms.
 To satisfy the renormalization condition \eq{Flow_condition} the heat kernel traces depend on $K(\phi)$ and thus have no off shell dependence on $\rho_1$ or $\rho_2$ such that the flow equation is of the form \eq{general_flow}.
 We note that strictly the heat kernel traces can only be evaluated when \eq{Kij} applies and therefore we cannot identify terms involving the trace free part of the extrinsic curvature  $K_{ij}^{\rm T} \equiv K_{ij} - \frac{1}{D-1} K \ga_{ij}$. However
 the first term in \eq{oneloopflow}  can be used to produce any term proportional to $K_{ij}^{\rm T}$ and this anyway does not affect the renormalization of Newton's constant.

Utilising the early-time heat kernel expansion on a manifold with a boundary \cite{mckean_jr.1967,Vassilevich:2003xt} we find the flow of the action $S$ with the boundaries present. Explicitly for $\mathcal{F}$ we find
\bea
  \Tr_2 [e^{- \Delta_2 }  ] - 2 \Tr_1 [e^{- \Delta_1 }  ]   &=&  \frac{ \frac{1}{2}D(D+1)  - 2 D \zeta_{\epsilon}^D }{(4 \pi )^{\frac{D}{2}}} \int d^Dx \sqrt{g} \nonumber   \\
  && +   \frac{1}{(4 \pi)^{\frac{D}{2}}}    \int_{\Sigma} d^{D-1}y \sqrt{\ga} \frac{\sqrt{\pi}}{2} \frac{1}{2} \left(D^2-4 (D-2) -3 D+4 \zeta _{\epsilon }^{D-1}\right)   \\
&&   +  \frac{1}{6} \frac{\frac{1}{2}D(D+1)  - 6 -  ( 2D + 12) \zeta_{\epsilon}^{D-2} }{(4 \pi )^{\frac{D}{2}}}  \left( \int d^Dx \sqrt{g} R   + 2  \int_{\Sigma} d^{D-1}y \sqrt{\ga} K \right)  + ... \nonumber
\eea
where we see that the required balance between the GHY term and the Einstein-Hilbert action is preserved. This result can be anticipated from the results of \cite{Jacobson:2013yqa} where it was shown that the required balance holds for the on shell Legendre effective action in $D=4$. There the result did not lead to a consistent picture since the balance needs to hold also off shell to identify the beta function. Here we see the required balance holds off shell. This is presumably the case since we have been careful not to break diffeomorphism invariance in deriving the RG equation \eq{oneloopflow}. 
Allowing for a field renormalization \eq{scaling_of_g_intro}, the balance is also preserved following the fact that both terms have the same canonical dimension. As such the beta function for Newton's constant is given by \eq{betas} derived either from the bulk or boundary action.

The renormalization of the boundary volumes is given by
\beq \label{beta_rho}
\beta_{\rho_\m} = (-D+1 + \eta_{V})  \rho_\m+    \frac{1}{(4 \pi)^{\frac{D}{2}}} \frac{\sqrt{\pi}}{2} \frac{1}{2} \left(D^2-4 (D-2) -3 D+4 \zeta _{\epsilon }^{D-1}\right) \,.
\eeq 
Let us note that we cannot put the constant term to zero if we also demand that the constant term for $\beta_{\lambda}$ is absent.
However, this is just a short coming of our regularisation scheme. We could add more parameters by normalising different components of the fields differently or by including matter fields (or even auxiliary fields) and adjusting their normalisation. Once this is done we can also remove the constant term from \eq{beta_rho} and will have a fixed point for $\rho_\m^*=0$. 

\section{The $\varepsilon$-expansion in quantum gravity}
\label{sec7}
As discussed in the introduction the $\varepsilon$-expansion in quantum gravity appears to give a fixed point for Newton's constant as an expansion in $\varepsilon$. However as first discussed in \cite{Kawai:1989yh} the loop-expansion is generically an expansion in $G/\varepsilon$ rather than in $G$. As such the  $\varepsilon$-expansion is not as one would naively expect.
 
To understand this first  recall that in two dimensions the Einstein-Hilbert action with a vanishing cosmological constant is a topological invariant.
 In consequence the theory in two dimensions is invariant under both diffeomorphisms and Weyl transformations $g_{\mu\nu} \to \Omega(x)^{-2} g_{\mu\nu}$. 
 Nonetheless when the limit $D \to 2$ is taken the topological nature of the Einstein-Hilbert action is evaded since the measure also becomes singular in this limit. This becomes clear if we canonically normalise the gauge invariant scalar degree of freedom. In particular  the hessian for the  gauge invariant scalar fluctuations of the metric $s$ is of the form (see \eq{shess}):
 \beq 
 S^{(2)}_{ss} = - \frac{1}{D^2} \frac{(D-2)(D-1)}{32 \pi G} \sqrt{g} ( - \nabla^2 + ... \,,
 \eeq
 which appears to vanish in the limit $D\to 2$, leading to a singular propagator.
 However the functional measure also involves the factor $-\frac{(D-2)(D-1)}{32 \pi G}$ and hence to perform the perturbative expansion one should canonically normalise $s$ by
 \beq \label{s_Normalise}
 s \to \sqrt{ - \frac{32  \pi G D^2}{(D-2)(D-1)}} s
 \eeq 
which removes the singular behaviour from the propagator. In consequence the vertices will have factors of $\sqrt{G/\varepsilon}$ not $\sqrt{G}$  and hence the perturbative expansion is really an expansion in 
\beq
G/\varepsilon \ll 1
\eeq
rather than in $G$.
Note that, since in the limit $D \to 2$ the field $s$ is the one gauge invariant degree of freedom, all other contributions to the renormalization originate from the measure. Consequently  there is no expansion in $G$ itself.

 Now the important point to realise is that if we impose that the theory should be Weyl invariant in the limit $D \to 2$ then we have a restriction on what we mean by an observable since, by definition,  an observable must be invariant. Thus one might suspect that using a physical scheme for which the reference observables are Weyl invariant in two dimensions will improve the situation.
 To achieve this one must include also matter fields $\psi$ where in two dimensions the reference observable $\mathcal{O}$ is invariant under
\beq \label{Weyl}
g_{\mu\nu} \to \Omega(x)^{-2} g_{\mu\nu} \,,\,\,\,\,\,\,\,\, \psi \to \Omega(x)^{d_{\psi} } \psi  \,,
\eeq    
where $d_{\psi}$ is the dimension of the field. Such an observable is provided by a four-fermion--$n_S$-scalar interaction since for fermions $d_{\psi} = (D-1)/2$  and $d_{\psi} = (D-2)/2$ for scalars. As we shall see, if we use a physical renormalization scheme where the reference observable is Weyl invariant in the limit $D \to 2$ no terms involving $G/\varepsilon$ are encountered and the beta function obtained at one-loop will already tell us where the fixed point $\beta(G_*)=0$ lies.

In this section we will investigate the fixed point near two dimensions and calculated the critical exponents.
We closely follow the previous work \cite{Kawai:1989yh,Kawai:1992fz} where dimensional regularisation was used. In these works it is pointed out that there occurs an over subtraction since the one-loop counter term for the conformal fluctuations is of order $O(\varepsilon^0)$ i.e.
\beq
\frac{(k)^\varepsilon }{\varepsilon} \sqrt{g}   R \sim (1  + \log( k) \varepsilon ) \partial_\mu s \partial^{\mu}s
\eeq  
where here $k$ is the IR renormalization scale. Thus one subtracts a finite term rather than a pole $1/\varepsilon$.  In \cite{Kawai:1992fz} a non-standard counter term was included in order to evade this perceived issue and in \cite{Kawai:1993mb} (and subsequent works \cite{Kawai:1995gt,Kawai:1995ju,Kawai:1996vt,Aida:1994dv,Aida:1994zc,Aida:1995ah,Aida:1996zn}) diffeomorphism invariance was sacrificed for the same reason. Here we do not perceive this as a problem since it is just a consequence of diffeomorphism invariance of action and the renormalization group invariance of the functional integral. Furthermore we want to renormalise gravity in higher dimensions where of course these are real divergencies.

 \subsection{Physical schemes and matter interactions near two dimensions}

In order to determine the beta function for Newton's constant near two dimensions and the scaling dimensions of various observables, we now consider a more general set of physical renormalization schemes based matter self interactions (or masses) 
\beq \label{O_of_L}
\mathcal{O}[g_{\mu\nu}, \psi] = \int d^Dx \sqrt{g} \mathcal{L}_{\rm int}(\psi)\,,
\eeq
which appears in the action with a coupling constant $\g$.  
If we denote by $d_{0}$ the classical scaling dimension of $\O$ then the scaling dimension will general receive an anomalous correction due to gravity  $d = d_{0}+ \eta$.\footnote{Here we do not include a subscript for the dimensions corresponding to $\O$ to avoid clutter; it should be understood that $d= d_{\O}$, $\g = \g_{\O}$ etc. } We then consider the action
\beq \label{actionD2}
S =   S_{EH}[g_{\mu\nu}] + S_{\psi}[g_{\mu\nu}, \psi]  + \g \int d^Dx \sqrt{g} \mathcal{L}_{\rm int}(\psi)
\eeq
where $S_{\psi}$ is the kinetic part of the matter field action which is conformally coupled:
the action is given by 
\beq
 S[\psi, g_{\mu\nu}] = \frac{1}{2} \int d^2x \sqrt{g}  \sum_{\n =1}^{N_S}   Z_{\n}^S \,  \psi_{S , \n} (-\nabla^2) \psi_{S,\n}  +  \sum_{\n =1}^{N_D} \, i  Z_{\n}^F \bar{\psi}_{F,\n} \Slash{\nabla} \psi_{F,\n }
\eeq
 with $\Slash{\nabla}$ denoting the Dirac operator. The central charge of the matter is given by $c_{\psi} = N_D + N_S$ where $N_D$ is the number of Dirac fermions and $N_S$ is the number of scalars. We will not consider boundaries in this section.

The case we have been studying up to this point is  $\mathcal{L}(g_{\mu\nu} ,\psi) = 1$ where $\mathcal{O} = \mathcal{V}$ and $\g = \lambda$. If we now consider a path integral with the interaction $\mathcal{O}$ 
instead of the cosmological constant term we can generalise our RG scheme. In particular we can consider the flow equation which takes the form \eq{flow_for_general_scheme} but where we impose 
\beq
-\partial_{t}  \langle \mathcal{O} \rangle  =   \left\langle d^a \frac{\delta}{\delta \phi^a} \mathcal{O} \right\rangle_{\g =0}
\eeq
where the field content $\phi= \{ g_{\mu\nu} , \psi \} $ now includes the matter fields $\psi$.
It follows that we should impose 
\beq
\frac{\partial}{\partial \g} \mathcal{F}  = 0  
\eeq  
on the coarse graining part of the flow equation. Furthermore we impose that 
the $Z_{\n}^F= 1= Z_{\n}^S $ such that no wave function renormalization of the matter sector is generated by gravity (we will neglect the renormalization of the matter interactions when $G=0$).
This can be achieved by introducing dimensionless matter fields 
\beq \label{phi_hat}
\psi = g^{-d_\psi/(2D)} \hat{\psi}
\eeq
such that
\beq
 \mathcal{L}(\psi) = \sqrt{g}^{-1-d_0/D }   \mathcal{L}(\hat{\psi}) 
\eeq
where $d_0 = -D(1- \Delta_0)$ and imposing that $\hat{\psi}$ has dimensions $d_{\hat{\psi}} =0$ also when quantum corrections are included along with the condition
\beq
\frac{\partial}{\partial Z_{\n}^F} \mathcal{F}  = 0  =  \frac{\partial}{\partial Z_{\n}^S} \mathcal{F}\,.
\eeq
In coordinates  $\phi =\{g_{\mu\nu} ,\hat{\psi}\}$ we have that 
\beq
d^a \frac{\delta}{\delta \phi^a}  =   d_g \,   \int d^Dx \,  g_{\mu\nu}(x) \frac{\delta}{\delta g_{\mu\nu}(x)}  \,,
\eeq 
and thus all scaling dimensions are encoded in the metric. 
We can then write down a one-loop flow equation close to two dimensions using the proper-time regulator. As we show in appendix~\ref{Hessians_in_two_dimensions}, the only modification to the flow equation near two dimensions 
is to replace $\mathcal{F}$ with   
\beq \label{F_general}
\mathcal{F} =  \Tr_2 [e^{- \Delta_2 }]   - 2 \Tr_1 [e^{- \Delta_1 }  ]  +   \Tr_0 [e^{-( -\nabla^2 -  \frac{d_0}{ 2 } R )} ] -  \Tr_0 [e^{-( -\nabla^2 + R)} ]  + {\rm matter \, contributions}
\eeq
where the extra terms follow from the modification of the way gravity couples to an operator of general dimension $d_{0}$.
This involves gauge invariant scalar $s$ which couples to the matter interactions which producing the third term. For $d_0 = -2$ the contribution from $s$ is cancelled by the fourth term which arises from the measure. When matter with an interaction $d_0 \neq -2$ is included this cancellation is no longer exact. The matter contributions will also lead to 
the renormalization of the Newton's constant as well as renormalization the matter couplings themselves. However the influence of gravity on matter is contained
in the anomalous dimension of the metric by the physical renormalization condition
\beq \label{etaO}
\eta_g(G)  = - \frac{2}{d_{0}}  \eta(G)   \,.
\eeq
One observes that for general $d_0$   keeping $\eta_g$ small would require  $\eta/d_0 \propto  G$  if the expansion was in $G$, on the other hand the expansion is in $\frac{G}{\varepsilon}$ and hence we expect $\eta/d_0 \propto  \frac{G}{\varepsilon}$.

\subsection{One-loop beta functions}

Within the schemes with reference observable of dimension $d_0$ we obtain the one-loop beta function
\beq \label{beta_general}
 \beta_G =  \varepsilon \left( 1  +   \frac{\eta}{d_0} \right) G  - \frac{2}{3}\left( 25 +  3 d_{0} - c_{\psi} \right) G^2  
\eeq
where the first term comes from first term of \eq{flow_for_general_scheme} and the second term comes from \eq{F_general}. 
The second term was first found in \cite{Kawai:1989yh} using dimensional regularisation where $\eta$ was set to zero. The beta function for the interaction couplings are given by
\beq
\beta_{\g} =  ( d_0 + \eta) \g   \,.
\eeq
Now we note that in any given scheme $\eta$ is unfixed by the beta functions. However if we fix the anomalous dimension $\eta$ in a single scheme  corresponding to a particular value of $d_0$ we will determine all over anomalous dimensions. Equivalently we can express the beta function for Newton's constant in the form
\beq
 \beta_G =  \varepsilon G  + \frac{2}{3}\left( c_g + c_{\psi} \right) G^2
\eeq    
where $c_g$ can be thought of as the central charge for the gravitational degrees of freedom.
Then comparing the two beta functions we arrive at the one-loop anomalous dimensions 
 given by
\beq
\eta = \frac{2}{3} d_0  \left(c_g+3 d_0+25\right)\frac{G}{\varepsilon }
\eeq
which depends on the number $c_g$ which remains undetermined. Comparing with \eq{etaO} we see that the anomalous dimension for the metric is small only if either $c_g = -3 d_0-25$ or $G/\varepsilon$ is small.  Since we know that in a generic scheme the expansion is in $\frac{G}{\varepsilon }$ this leaves the value of $c_g$ undetermined without further insight.

\subsection{Higher loops and the UV fixed point}
\label{higherloops}

Now if we were to go to higher loop orders the beta function for Newton's constant will be given by
\beq \label{beta_series}
 \beta_G = \varepsilon \left( 1  -   \frac{\eta_g}{2} \right) G  - \frac{2}{3}\left( 25 +  3 d_{0} - c_{\psi} \right) G^2      + G^2 (  b_2(d_0) \frac{G}{\varepsilon} + b_3(d_0) \frac{G^2}{\varepsilon^2} + ...)
\eeq  
where the coefficients $b_\n$ will depend on $d_0$. Let us now consider the case where the reference observable is Weyl invariant in two dimensions which means that 
\beq
d_0 = a  \varepsilon + O(\varepsilon^2)
\eeq
for some constant $a$. In this case all of the coefficients $b_\n$ will vanish for $\varepsilon \to 0$. This is seen most easily by exploiting the dimensionless parameterisation of the matter fields \eq{phi_hat} and using the conformal gauge for the gravitational degrees of freedom
\beq \label{exponential_conformal_gauge}
g_{\mu\nu} = e^{2  \sqrt{ - 8 \pi G/\varepsilon} \sigma} \hat{g}_{\mu\nu} 
\eeq
where $\hat{g}_{\mu\nu}$ is gauge fixed up to topological fluctuations. In this parameterisation the Einstein-Hilbert action becomes the canonically normalised Liouville action\footnote{More generally the two-dimensional  limit of the Einstein-Hilbert action with $G \sim \varepsilon$ is related to the covariant Polyakov action \cite{Nink:2015lmq} which reduces to the Liouville action in the conformal gauge }:
\bea
S[\phi]&=& - \frac{1}{16 \pi G} \int d^Dx \sqrt{\hat{g}} \left[ \hat{R} \left( 1 + \varepsilon  \sqrt{ - 8 \pi G/\varepsilon}   \sigma \right)   - 8 \pi G  \hat{g}^{\mu\nu} \partial_{\mu} \sigma \partial_{\nu} \sigma \right] \nonumber  \\
&+& \g \int d^Dx \sqrt{\hat{g}}  \mathcal{L}(\hat{\psi}) + S_{\psi}[\hat{g}_{\mu\nu}, \hat{\psi}]  + O(\varepsilon) \,.
\eea
As such the integral over the gravitational degrees of freedom becomes 
gaussian\footnote{ Strictly the measure makes the integral non-gaussian but the ``vertices'' which enter at two-loops and beyond lead only to non-universal divergencies which are set to 
zero using dimensional regularisation.}, while decoupling from the Weyl invariant reference observable.  Under the reasonable assumption that all to loop orders the the evaluation of the functional integral is independent of the parameterisation of 
the physical degrees of freedom, (remember there is no source term present) this is just a convenient choice of coordinates. 
The important point is that by using Weyl invariant reference observables to define the renormalization scheme, we guarantee that the higher-loop coefficients in \eq{beta_series} vanish.

 In this case the loop expansion is no longer an expansion in $G/\varepsilon$ and hence to keep the anomalous dimension of the metric small we require that $\eta_g \propto G$. As a result the beta function for Newton's constant is given by
\beq \label{beta_weyl}
 \beta_G = (D-2) G  - \frac{2}{3}\left( 25  - c_{\psi} \right) G^2  
\eeq 
with all higher loop terms being zero in the limit $\varepsilon \to 0$.
This beta function agrees with the beta function computed in exactly two dimensions using Liouville theory \cite{Polyakov:1981rd,Distler:1988jt,David:1988hj,Codello:2014wfa}.   
 From \eq{beta_weyl} we see that there exists a UV fixed point at
\beq \label{Gstar}
G_* = \frac{3}{2} \frac{D-2}{(25 - c_{\psi})}  \,,
\eeq
where we observe that $G_*$ is positive for $N_S+ N_D < 25$.
Although the fixed point \eq{Gstar} has been found previously \cite{Kawai:1989yh}, here we observe that \eq{Gstar} is not an approximation. In particular it is exact when we exploit dimensional regularisation which sets all non-universal terms in $\beta_G - \varepsilon G$, that  vanish for $\varepsilon \to 0$, to zero. As such the fixed point exists for all dimensions $D>2$ since the $\varepsilon$-expansion for the fixed point only has a linear term.

The running $G$ in the weakly coupled phase $0\leq G \leq G_*$ can be expressed in terms of the RG time $t$ with the reference scale taken to be the IR Planck mass $\Lambda_0 = M_{\rm Pl}>0$ as
\beq
G(t) = \frac{ e^{t (D-2)}}{ 1 + 1/G_* e^{t(D-2)}} \,.
\eeq
Where for all values of $M_{\rm Pl} >0$ we run from the UV fixed point to the IR fixed point where the dimensionful Newtons constant is given by $M_{\rm Pl}^{-2}$.

It should be remarked that a consequence of the one-loop exactness of the beta function implies that the Einstein-Hilbert action scales canonically
\beq
- \partial_{t}  \left\langle \int d^Dx \sqrt{g} R \right\rangle = - (D-2) \left\langle \int d^Dx \sqrt{g} R \right\rangle 
\eeq
and thus it is possible to consider the Einstein-Hilbert action itself as the reference observable. 
This would not be the case if the higher loops did not vanish.  Since   $ \int d^Dx \sqrt{g} R$ vanishes on-shell and appears as an eigen-perturbation of the fixed point it is seen to be a redundant operator at the UV fixed point \cite{Dietz:2013sba}.
This reflects the fact that the fixed point action itself is not scheme independent  and we have just chosen the natural scheme, i.e. dimensional regularisation, where the action is of the Einstein-Hilbert form.
In other schemes the action can take a different form even though the universal critical exponents should be the same.

\subsection{One-loop anomalous dimensions}
Let us stress that although we exploited a particular scheme to obtain the exact beta function scheme independence is restored as the exact level.   
In particular the exact one-loop beta functions can be made to agree and thus by comparing the expression for the beta functions  \eq{beta_weyl} and \eq{beta_series} in different schemes  we can determine the anomalous for observables with $d_0 \sim O(1)$. Comparing the one-loop expressions  \eq{beta_general} and \eq{beta_weyl} we can then infer that the one-loop anomalous dimensions for an observable of dimension $d_0$ are given by
 \beq \label{eta_one_loop}
 \eta = 2 d_0^2 \frac{G}{\varepsilon} + O(G^2/\varepsilon^2) 
 \eeq
and that at the fixed point \eq{Gstar} we obtain
\beq \label{etastar_one_loop}
\eta_* = 3 d_0^2 \frac{1}{25 - c_{\psi}}+ O(1/(25 - c_{\psi})^2)\,.
\eeq
However, since this calculation involves breaking Weyl invariance we have to resum the loop-expansion in $G/\varepsilon$ in order to obtain the leading order critical exponents in the $\varepsilon$-expansion.

\subsection{Two-dimensional quantum gravity}
\label{2Dquantumgravity}

An important question is whether two-dimensional gravity can be obtained from the $\varepsilon \to 0$ limit.
In \cite{Kawai:1992fz} it was suggested that this is achieved by setting $G = - G_*$ and then taking the limit.
In fact there is a good reason for this since $G = -G_*$ is nothing but the IR fixed point in two dimensions when $c_{\psi} < 25$.
Let us now explain how this comes about.

First we observe that the two-dimensional beta function for Newton's coupling is given by
\beq
\beta_G = - \frac{2}{3} (25 - c_{\psi}) G^2
\eeq
which for $c_{\psi} < 25$ has a UV fixed point at $G \to + 0$ and an IR fixed point at $G \to - 0$. On the other hand for $D>2$ there is a UV fixed point \eq{Gstar}
which goes to $G \to + 0$ in the limit $\varepsilon \to 0$. Now the key point to realise is that, from the two-dimensional point of view, $\varepsilon > 0$ plays the role of an IR cutoff
within dimensional regularisation. Thus the bare Newton's coupling is given by
\beq
\frac{1}{G}   =  - \frac{1}{\varepsilon}  \frac{2(25 - c_{\psi})}{3}  
\eeq
 such that the IR divergence occurs for $\varepsilon \to 0$. Thus one observes that the bare coupling is sent  to $G \to - G_*$ thus we find that the IR fixed point in  $D=2$ is at
\beq
G_{\rm IR2D} = - G_*
\eeq
in the limit $\varepsilon \to 0$. Note that this is also a fixed point of \eq{beta_weyl} but only when the two-dimensional limit is taken.
This suggests that we evaluate $\eta$ at $G = G_{\rm IR2D}$ to obtain the scaling exponents in two-dimensional quantum gravity.
As we showed in section~\ref{two_D_limit_measure} the fact that $G_{\rm IR2D} \propto \varepsilon$ means that the measure in this limit is not singular.
On the other hand if we keep $G$ fixed and take $D\to2$ the measure will be singular.

\subsection{Non-perturbative calculation}
\label{Scalaing_exponents_calculation}

We now wish to calculate the anomalous scaling dimensions $\eta$ for observables with a non-vanishing classical dimension $d_0$ in the limit $D \to 2$ while keeping $G/(D-2)$ constant. 
This holds at the fixed points $G=G_*$ and $G= G_{\rm  IR2D}$ where the scaling dimensions  correspond to the scaling exponents at the UV fixed point in $D> 2$ dimensions
and at the IR fixed point in two dimensions respectively. Since for $d_0 \neq0$ we break Weyl invariance the loop expansion is an expansion in $G/\varepsilon$ and  to obtain 
non-perturbative critical exponents one must resum this series. On the other hand the critical exponents in two dimensions are known exactly and are given by \cite{Distler:1988jt,David:1988hj}\footnote{See \cite{Codello:2014wfa} for a calculation of these exponents using the functional renormalization group.}
\beq \label{beta_exact}
d_{\rm  IR2D}  \equiv -2 \beta  = -2 (25 -c_{\psi})  \frac{1 - \sqrt{1 + 24 \frac{1}{(25- c_{\psi})}   (\Delta_0 -1)}}{12 }
\eeq 
where here we use the standard notation in two dimensions
\beq \label{Delta0}
d_0(D=2) \equiv -2(1-\Delta_0) 
\eeq
for the classical dimensionality of the observable. Equivalently by denoting the scaling dimension of the volume by $\alpha \equiv  \beta|_{\Delta_0=0}$ the
relative scaling dimension $\Delta \equiv 1- \beta/\alpha$ satisfies the KPZ relation \cite{Knizhnik:1988ak}:
\beq
\Delta - \Delta_0 = \frac{6 \alpha^2}{25 - c_{\psi}} \Delta ( 1- \Delta)  \,.
\eeq
If we now take the one-loop approximation we evaluate \eq{eta_one_loop} at $G= G_{\rm IR2D}$
to obtain 
\beq
-2 \beta = -2(1-\Delta_0) -12 \frac{(1-\Delta_0)^2}{25 - c_{\psi}}  + O( 1/(25 - c_{\psi})^2 )
\eeq
which agrees with the exact result to this order. In \cite{Kawai:1992fz} it was shown that the exact critical exponents \eq{beta_exact} can be obtained by re-summing the loop expansion for
$G=G_{\rm 2DIR}$. However it is straightforward to perform the same calculation for general $G/\varepsilon \sim O(1)$ and therefore to obtain the critical exponents at the UV fixed point 
$G =G_*$. Since in dimensional regularisation all quantum corrections will be evaluated in two dimensions we will use  $\Delta_0$ defined by \eq{Delta0} to express this two dimensional classical scaling dimension and $d_0$ for the $D$-dimensional classical scaling dimension which we retain only at tree-level.

To perform this calculation we make we again make use of a particular form of the  the conformal gauge such that Einsteins theory is a free theory close to two dimensions.   
This can by achieved by first writing \cite{Kawai:1992fz}
\beq
g_{\mu\nu} =  \left( 1 + \frac{\varepsilon}{2} \sigma \right)^{\frac{4}{\varepsilon}} \hat{g}_{\mu\nu}
\eeq 
where $\hat{g}_{\mu\nu}$ is a metric with unit determinant which we then gauge fix. This parameterisation takes the form of an exponential in the limit $\varepsilon \to 0$.
In terms of the field variables $\hat{g}_{\mu\nu}$ and $\sigma$ the Einstein-Hilbert action is given by
\beq
- \frac{1}{16 \pi G} \int d^Dx \sqrt{g} R = - \frac{1}{16 \pi G} \int d^Dx \sqrt{\hat{g}} \left[ \hat{R} \left( 1 + \frac{\varepsilon}{2} \sigma \right)^2  + (D-2)(D-1) \hat{g}^{\mu\nu} \partial_{\mu} \sigma \partial_{\nu} \sigma \right] \,,
\eeq
where up to topological fluctuations $\hat{g}_{\mu\nu}$ is pure gauge in the limit $D \to 2$.
 To canonically normalise $\sigma$, which plays the role of the gauge invariant scalar $s$, we perform the replacement
\beq
\sigma \to \sqrt{ -  \frac{8 \pi G}{(D-2)(D-1)}} \sigma
\eeq
which removes also these factors from the functional measure. Removing the pole from the propagator for $\sigma$ when  $D \to 2$. In the limit $D\to2$ we would then recover \eq{exponential_conformal_gauge}.  Around flat spacetime $\hat{g}_{\mu\nu}= \eta_{\mu\nu}$ one has just a canonically normalised scalar field 
\beq
- \frac{1}{16 \pi G} \int d^Dx \sqrt{g} R =   \frac{1}{2} \int d^Dx \sqrt{\hat{g}}  \hat{g}^{\mu\nu} \partial_{\mu} \sigma \partial_{\nu} \sigma  \,,
\eeq
and thus the theory is free which makes the perturbative treatment straight forward.
The propagator for the mode $\sigma$ around flat spacetime is then just
\bea
 \mathfrak{G}(p^2) =  \frac{1}{p^2}
\eea
and thus when performing the loop expansion each momentum integral will be regularised to obtain
\beq
\int \frac{d^Dp}{ (2 \pi)^D }   \mathfrak{G}(p^2)  =  - \frac{1}{2 \pi}  \frac{k^{\varepsilon}}{\varepsilon} + O(\varepsilon^0)
\eeq
by dimensional regularisation with $k$ the IR renormalization scale. 
It follows that we can write down a zero dimensional propagator 
\beq
\mathfrak{G} = - \frac{1}{2 \pi}  \frac{k^{\varepsilon}}{\varepsilon}  \,,
\eeq 
which then appears in place of the standard Feynman rule.
Then the functional integral for the conformal factor is reduced to 
\beq \label{Zsigma}
\mathcal{Z}_{\sigma}(k) = \mathcal{N} \int_{-\infty}^{\infty} d (i \sigma)  e^{ \pi k^{-\varepsilon} \varepsilon \sigma^2 } 
\eeq
where here we are working in units of the UV scale $\Lambda$ and we note that in fact the we should reverse the Wick rotation of $\sigma$ by sending $\sigma \to - i \sigma$ such that the Gaussian integrals have the right sign.
To normalise the functional integral we should take $\mathcal{Z}_{\sigma}(k)|_{\varepsilon \to 0} =1$ which determines that $ \mathcal{N} = \sqrt{\varepsilon}$.

Now to calculate the averages of observables \eq{O_of_L}, which in terms of the dimensionless fields \eq{phi_hat} take the form of composite operators \footnote{See \cite{Pagani:2016dof} for a study of composite operators in quantum gravity using the functional renormalization group.} ,
\beq
\mathcal{O} = \int d^Dx \sqrt{g}^{-\frac{d_0}{D}} \mathcal{L}(\hat{\psi}) \,,
\eeq
we can use the standard integral \eq{Zsigma} where all fields now live in zero-dimensions.
In particular the averages are given by
\beq
\langle \mathcal{O}  \rangle_k = \frac{\sqrt{\varepsilon}}{\mathcal{Z}_{\sigma}(k)} \int d^Dx  \sqrt{\hat{g}}^{-\frac{d_0}{D}}  \mathcal{L}(\hat{\psi})   \int_{-\infty}^{\infty} d\sigma   \left( 1 +  \sqrt{  \frac{8 \pi G}{\varepsilon}} \frac{\varepsilon}{2} \sigma \right)^{\frac{4}{\varepsilon} (1- \Delta_0)  }     e^{- \pi k^{-\varepsilon} \varepsilon \sigma^2 }  
\eeq
where now if we expand in $G/\varepsilon$ we will produce the loop expansion we want to resum. In particular we want to take the limit $\varepsilon \to 0$ while avoiding the expansion in $G/\varepsilon$. To do so one then makes the change of variables 
\beq
\sigma \to \sigma/\varepsilon
\eeq
in order that we can apply the method of steepest descent where $\varepsilon$ is the small parameter.
 The integral is given by
\beq
\langle \mathcal{O}  \rangle_k =  \frac{1}{\sqrt{\varepsilon} \mathcal{Z}_{\sigma}(k)}  \int d^Dx \sqrt{\hat{g}}^{-\frac{d_0}{D}}  \mathcal{L}(\hat{\psi})   \int_{-\infty}^{\infty} d\sigma     \exp \left\{   \frac{1}{\varepsilon} \left( 4 (1- \Delta_0)   \log  \left( 1 +  \sqrt{  \frac{8 \pi G}{\varepsilon}} \frac{1}{2} \sigma \right)  - \pi k^{-\varepsilon}  \sigma^2 \right) \right\}  
\eeq
Let us note that after performing all redefinitions of $\sigma$ we have
\beq
g_{\mu\nu} =  \left( 1 + \frac{1}{2}  \sqrt{ \frac{8 \pi G}{(D-2)(D-1)}}   \sigma \right)^{\frac{4}{\varepsilon}} \hat{g}_{\mu\nu}
\eeq 
which is a parameterisation which sets up an $\varepsilon$ expansion i.e it ensures that the action is quadratic in the field and proportional to $1/\varepsilon$.
We can make a saddle point approximation by writing
\beq
\sigma = \sigma_0 + \sqrt{\varepsilon} \delta \sigma
\eeq
where inside $ \mathcal{Z}_{\sigma}(k)$ we have $\sigma_0 =0$ and inside the integral over $\mathcal{O}$ the saddle point 
$\sigma_0$ should minimises the `potential'
 \beq
\frac{\partial}{\partial \sigma} \left( k^{-\varepsilon} \pi  \sigma^2 -  4 (1 - \Delta_0) \log ( 1 + \sqrt{ 2 \pi G/ \varepsilon }  \sigma )  \right) = 0 \,
\eeq
which has two solutions 
\beq \label{sigma0}
\sigma_0 = - \frac{ 1  \pm \sqrt{ 1 - 16 G \varepsilon^{-1} k^{\varepsilon} (1- \Delta_0)}}{2 \sqrt{2 \pi} \sqrt{ G \varepsilon^{-1} } }\,.
\eeq
Performing the saddle point approximation we then have the expression
\beq 
\langle \mathcal{O}  \rangle_k \approx  \int d^Dx  \sqrt{\hat{g}}^{-\frac{d_0}{D}}  \mathcal{L}(\hat{\psi})   \exp \left\{   \frac{1}{\varepsilon} \left( 4 (1- \Delta_0)   \log  \left( 1 +  \sqrt{  \frac{8 \pi G}{\varepsilon}} \frac{1}{2} \sigma_0 \right)  - \pi k^{-\varepsilon}  \sigma^2_0 \right) \right\}  
\eeq
from which we can extract the anomalous dimensions.
Since here $k$ is the IR cutoff the anomalous dimension can be obtain by
\beq \label{dkO}
k \partial_k \langle \mathcal{O}  \rangle = \eta  \langle \mathcal{O}  \rangle \,.
\eeq
Equally we may take a derivative with respect to the UV cutoff scale $\Lambda$. In this case we should use the scaling laws
$- \partial_t \hat{g}_{\mu\nu} = -2  \hat{g}_{\mu\nu} $, $\partial_t \sigma  = 0 = \partial_t \hat{\psi}$  for the fields,
$- \partial_t k = k$ for the IR renormalization scale and $\partial_t G = \beta_G$ for Newton's constant.  Then we obtain the scaling dimension by the familiar expression 
\beq \label{dLambdaO}
- \partial_t \langle \mathcal{O}  \rangle =  d  \langle \mathcal{O}  \rangle
\eeq
with $d =  d_{0} + \eta$ agreeing with \eq{dkO} provided we are at a fixed point $\beta_G =0$. 
Using either \eq{dkO} or \eq{dLambdaO}  yields the scaling dimension given by
\beq \label{d_non_perturbative}
d(G) =  d_0(D) + \frac{1 - \sqrt{1 - 16 \frac{G}{D-2}  (\Delta_0 -1)}}{4  \frac{G}{D-2} } + 2(1-\Delta_0) 
\eeq
where we choose the negative root solution \eq{sigma0} such that for $G/\varepsilon \to 0$ we recover the $d \to d_0$ .

To obtain the critical exponents $\beta$ at the IR fixed point in two dimensions we take $G =  G_{2DIR}$ and then take $\varepsilon \to 0$ recovering exact result \eq{beta_exact}. 
 On the other hand if we set $G = G_*$ we obtain the critical exponents at the UV fixed point \eq{Gstar} defined by $\theta \equiv - d_*$ which are given by
  \beq \label{theta}
  \theta = - \frac{1}{6} (25-c_{\psi}) \left(1-\sqrt{1-\frac{12 d_{0}}{25-c_{\psi}}}\right) - 2(1-\Delta_0) - d_0 \,.
  \eeq
  If we expand in $1/(25- c_{\psi})$ we recover the one-loop result \eq{etastar_one_loop}.
  
  Away from the fixed point the running of the couplings $\g$ is given by
  \beq \label{beta_g_exact}
  \beta_\g =  d(G) \, \g   \,,
  \eeq
which is non-perturbative in $G$.
  
An interesting outcome of this prediction is that, although the critical exponents of two-dimensional quantum gravity at the IR fixed point and the critical exponents at the UV fixed point in higher dimensions differ as $D\to2$, they are nonetheless related by analytical continuation $G_* \to -G_*$.   
 The theories obtained in different limits for quantum gravity close to two dimensions are summarised in 
table~\ref{table}. 

\begin{table}[h]
\begin{center}
\scalebox{0.88}{
\begin{tabular}{|c|c|c||}
\hline
Newton's Constant & Dimension & Theory \\
\hline
$G \to 0$ & $D> 2$ & Classical gravity  in $D>2$ dimensions   \\
\hline
$G \to G_*$ & $D> 0$ & Continuum limit of quantum gravity in  $D>2$ dimensions   \\
\hline
 $G \neq 0$ & $D \to 2$ & Singular  \\
 \hline
$G = -G_*$  & $D\to 2$  &  Two-dimensional quantum gravity  \\
\hline
\end{tabular}
}
\end{center}
\caption{ \label{table} The table shows the which theories the various phases of quantum gravity in $D > 2$ dimensions correspond to.  
In dimensions higher than two there is an IR fixed point where the Newton's constant vanishes. The continuum limit of this theory is taken at the UV fixed point where $G_*$ is finite.
If one starts in higher dimensions and takes the limit to two dimensions the functional integral becomes singular. However if we first go to the $G = -G_*$ and then take the limit $D \to 2$ we recover IR fixed point of two-dimensional quantum gravity.}
\label{t1}
\end{table}

\subsection{Non-perturbative scheme independence}
So far we have identified the fixed point for Newton's constant based on the physical scheme which preserved two-dimensional Weyl invariance.
On the other hand universal results should not depend on this choice which is just a scheme allowing us to compute the non-perturbative beta function with ease.
With the non-perturbative beta functions at hand let us now write out the full beta function for Newton's constant in an general physical scheme.
We write first that the exact beta function in a general scheme is given by
\beq
\beta_G^{\rm exact} = \varepsilon\, G - \frac{\eta_g}{2} \varepsilon G  + \tilde{\beta}(G) 
\eeq
where we determine $\tilde{\beta}(G)$ for a scheme based on an observables with dimension $d_0 = -2(1 - \Delta_0)$ in two dimensions by comparing to the beta function obtained in the Weyl invariant scheme and using the physical renormalization condition \eq{etaO}. This then leads to the identity 
\beq
 - \frac{\varepsilon G}{2 (\Delta_0 -1)}  \left(  \frac{1 - \sqrt{1 - 16 \frac{G}{\varepsilon}  (\Delta_0 -1)}}{4  \frac{G}{\varepsilon} } + 2(1-\Delta_0) \right)+ \tilde{\beta}(G)  =   - \frac{2}{3} (25 - c_\psi) G^2 \,.
\eeq
Thus for a general scheme the exact beta functions is given by
\beq
\beta_G= \varepsilon \, G     - \frac{2}{3} (25 - c_\psi) G^2    - \frac{\eta_g}{2} \varepsilon \,G       - \frac{\varepsilon G}{2 (\Delta_0 -1)}  \left(  \frac{1 - \sqrt{1 - 16 \frac{G}{\varepsilon}  (\Delta_0 -1)}}{4  \frac{G}{\varepsilon} } + 2(1-\Delta_0) \right) \,.
\eeq
Now if we were to expand in $G$ we would get the loop expansion 
\bea
\beta_G &=& \varepsilon \,G   - \frac{\eta_g}{2} \varepsilon \, G    +  \frac{2}{3} G^2 \left(c_{\psi }-6 \Delta _0-19\right)-\frac{32 \left(\Delta
   _0-1\right){}^2 G^3}{\varepsilon }-\frac{320 \left(\Delta _0-1\right){}^3
   G^4}{\varepsilon ^2} \\
   &-& \frac{3584 \left(\Delta _0-1\right){}^4 G^5}{\varepsilon
   ^3}-\frac{43008 \left(\Delta _0-1\right){}^5 G^6}{\varepsilon ^4}-\frac{540672
   \left(\Delta _0-1\right){}^6 G^7}{\varepsilon ^5}  \nonumber \\ 
   &-&\frac{7028736 \left(\Delta
   _0-1\right){}^7 G^8}{\varepsilon ^6}+O\left(G^9\right) \nonumber
\eea
and come to the conclusion that taking the limit $\varepsilon \to 0$ was not possible.
However this is only an artefact of perturbation theory. If we instead take the limit $\varepsilon$ for the exact expression we have 
\beq
\beta_G(\varepsilon \to 0) =  - \frac{2}{3} (25 - c_\psi) G^2 \,.
\eeq 
It then follows that within dimensional regularisation the exact beta function in dimensions $D>2$ is given by \eq{beta_weyl} independently of the renormalization scheme.
 

\subsection{Non-perturbative renormalization}
At the asymptotically safe fixed point $G=G_*$ an observable is relevant if the real part of the exponent $\theta$ is positive, $\Re (\theta) > 0$ whereas for $\Re (\theta) < 0$  the corresponding operator is irrelevant and the fixed point predicts that $\g = 0$.
For an  $n_S$-scalar--$n_F$-fermion interaction:
\beq
\mathcal{L}_{\rm int}(\psi) = \psi_S^{n_S} (\bar{\psi}_F \psi_F)^{\frac{n_F}{2}}
\eeq
 we have $ d_0(D) = -D + n_S (D-2)/2 + n_F (D-1)/2$ which gives
\bea
\theta =   -\frac{1}{6} (25-c_{\psi}) \left(1-\sqrt{1-\frac{12
  \left(\frac{n_F}{2}-2\right)}{25-c_{\psi}}}\right)    + \frac{1}{2} (D-2)  (2-n_F- n_S ) \,.
  \eea  
We observe that for all $c_{\psi}$ there is always a finite number of relevant interactions in integer dimensions $D>2$ since the real part of the first term is bounded whereas the second term, which is  proportional to $D-2$, decreases as the number of powers of the fields in the interactions increases.

\section{Discussion}
 \label{conclude}

In this paper we have sought to carefully refine the application of the renormalization group to gravity in order to study the asymptotic safety
by means of the $\varepsilon$-expansion.
This is motivated by the problem that beta functions can appear to depend on the parameterisation of physical 
 degrees of freedom. The dependence is understood more generally as a dependence on the 
 renormalization scheme and can be compensated by a renormalization of the fields. Since neither the 
 parameterisation of the fields, nor the renormalization of the fields, is physical, this suggests we take a different approach. Here we have defined 
 physical renormalization schemes where the unphysical dependencies are replaced by the dependence on
  scaling dimensions of physical observables.

  Working directly with physical observables, rather than local 
  correlation functions, also a great technical convenience since the equations become reparameterisation invariant. As such
  one can use the choice of parameterisation and gauge fixing to one's advantage, i.e. to simplify the problem at hand, safe in the knowledge that one is
  not implicitly modifying the renormalization scheme.  Of course this hinges on the regularisation scheme being reparameterisation and diffeomorphism invariant.
   At the classical level, diffeomorphism invariant and background independent flow equations have 
  been derived in \cite{Morris:2016nda}. Here the flow equations we have used achieve this already at one-loop (notwithstanding the issue of finding suitable boundary conditions), however the proper-time regularisation breaks down at the two-loop level. As such we have then used dimensional regularisation to achieve a non-perturbative result.
  As advocated in \cite{Morris:2016nda} constructing an exact diffeomorphism invariant flow equation could be achieved by using supersymmetric Pauli-Villars fields.
  It would also be desirable if such an equation was reparameterisation invariant.

 Here we have seen that adopting a reparameterisation, diffeomorphism and background independent approach bears many fruits.
Exploiting dimensional regularisation a UV fixed point can be identified since the non-perturbative beta function in the limit $D \to 2$ is just given by the conformal anomaly.
It remains to see if only a finite number of interactions are relevant. Here we have considered just interactions involving fermions and scalars 
 finding that this requirement is fulfilled. One should also include higher orders in derivatives and gauge fields to see if this picture persists.
 While the fixed point is Gaussian in the matter sector used here, we cannot include free gauge fields since they break two-dimensional Weyl invariance.
 This suggests that the fixed point for gauge fields is non-trivial.
 We also need to construct the renormalisable trajectories that move away 
 from the UV fixed point, towards low energies, to see whether the predictions of general relativity and the standard model can be reproduced. Since here we observe nothing special about four dimensions this leaves open the possibility of extra dimensions.

  It should be duly noted that there are strong parallels between the asymptotic safety scenario we uncover here and non-critical string theory in $D = c_{\psi} +1$ dimensions.
  In addition, the fact that the critical exponents are obtained almost directly from two-dimensional quantum gravity indicates that the fractal dimension of spacetime may be close to two. This observation was first made in causal dynamical triangulation simulations \cite{Ambjorn:2005db} and has since also been observed in other approaches to quantum gravity \cite{Lauscher:2005qz,Modesto:2008jz,Calcagni:2014cza}. 
  Taking the radically conservative view that Nature is indifferent to how we parameterise her, it could be the case that quantum gravity is described both by 
  string theory and a genuine non-perturbative quantisation of general relativity.   
  
\section*{Acknowledgements}
I would like to thank   Stanley Deser, Michael Duff, Mikhail Kalmykov, and  Aron Wall for constructive comments on \cite{Falls:2015qga}. This work has benefited from discussions with Dario Benedetti, Alessandro Codello, Nicolai Christiansen, Astrid Eichhorn, John Gracey, Stefan Lippoldt, Tim Morris, Carlo Pagani,  Jan Pawlowski and  Masatoshi Yamada. I also thank Alejandro Satz for correspondence on the diffeomorphism invariant boundary conditions. 
This work was supported by the  European Research Council grant ERC-AdG-290623.

\begin{appendix}

\section{Factoring out the gauge modes}
\label{A1}
Here we take a geometrical approach to the functional integral  viewing the fields $\phi^a$ as coordinates on a manifold $\Phi$ which can be thought of as a product of the physical space $\Phi/\mathcal{G}$ and the gauge orbits $\mathcal{G}$ with coordinates $\xi^{\alpha}$. As an alternative to the usual Faddeev-Popov gauge fixing we take the geometrical approach to factoring out the gauge modes \cite{Mazur:1989by,Bern:1990bh, Mottola:1995sj,Vassilevich:1995bg}) which keeps gauge independence manifest.
First we take the measure over the fluctuation fields $\delta\phi^{a}$
\beq
\int \prod_a   \frac{d\phi^a}{(2\pi)^{1/2}}  \sqrt{\det C_{ab}} =  \int \prod_a   \frac{d\delta\phi^a}{(2\pi)^{1/2}}  \sqrt{\det C_{ab}} 
\eeq
and decompose the fluctuation as:
\beq \label{decompose}
\delta \phi^a = L^{a}_{\bar{a}}  f^{\bar{a}} + L^a_{\alpha}  \xi'^{\alpha}
\eeq
where the second term is a diffeomorphisms with $\xi^{\alpha}$ parameterising the gauge orbit and   $f^{\bar{a}}$
are gauge invariant fields. The prime here indicates that zero modes of $L^a_{\alpha}$ must be left out of the spectrum of $\xi'$ which for gravity corresponds to Killing vectors.
In the case of gravity it is not possible to diagonalise the field space metric in these coordinates, however one can make an additional shift, corresponding to the freedom to fix the gauge
\beq \label{shift}
f^{\bar{a}} \to f^{\bar{a}} + t^{\bar{a}}_{\alpha} \xi'^{\alpha}  \,, \,\,\,\,\,\,   \xi'^{\alpha}  \to \xi'^{\alpha}
\eeq
which has unit Jacobian and hence does not alter the measure. Choosing $t^a_{\alpha}$  the DeWitt metric can be made block diagonal:
\beq \label{decompmetric}
 \delta\phi^a \, C_{ab}\, \delta\phi^b  =   f^{\bar{a}} C_{\bar{a} \bar{b} } f^{\bar{b}} + \xi'^\beta C_{\alpha \beta} \xi'^{\alpha}
\eeq
where here $\bar{a} ,\bar{b},..$ are a set of DeWitt indices $\bar{a} = \{ x, \bar{A}\}$ for the gauge invariant fields and $\alpha,\beta,...$ is a set of DeWitt indices for the diffeomorphisms e.g $\xi^\alpha = \epsilon^{\mu}(x)$.  
Next we write the gauge volume as
\beq
 V_{\rm diff}=  \int  \prod_{\alpha} \frac{d\xi^{\alpha}}{(2 \pi)^{1/2}} \sqrt{\det G_{\alpha \beta}}  \equiv  \Omega \int  \prod_{\alpha} \frac{d\xi'^{\alpha}}{(2 \pi)^{1/2}} \sqrt{{\det}' G_{\alpha \beta}}
\eeq
which comes with its own metric $G_{\alpha \beta}$. Here $\Omega$ is the volume of the subgroup of diffeomorphisms $\mathcal{H}$ which are zero modes of $ L^a_{\alpha}$ and the prime indicates that these modes are removed from the determinant. 
The total measure is then given by
 \beq
\frac{\int \prod_a d\delta\phi^a \sqrt{\det C_{ab}} }{ V_{\rm diff} }  =  \frac{1}{\Omega}  \int \prod_{\bar{a}} \frac{d f^{\bar{a}}}{(2\pi)^{1/2}}  \sqrt{{\det}' (G^{-1})^{\alpha \beta}}  \sqrt{{\det}' C_{a b}}   \,.
\eeq
where all gauge modes have been factored out apart from the zero modes that must be accounted from by determining $\Omega$ explicitly \cite{Volkov:2000ih} (a similar factor is needed for Maxwell theory \cite{Donnelly:2013tia}). 

 It is important to bare in mind that different choice of the fundamental degrees of freedom $\phi^A(x)$ can lead to unphysical configuration spaces of different dimensionality. For example if we choose the metric $\phi^A(x)=  g_{\mu\nu}$ or the Dirac matrices $\tilde{\phi}^{\tilde{A}}(x) =  \gamma_\mu$ then the number of `flavours', i.e values $A$ and $\tilde{A}$ can take, is different but the same is true for the gauge orbits parameterised by $\xi^\alpha$ and $\tilde{\xi}^{\tilde{\alpha}}$.    
If the physical degrees of freedom are the same then one expects that two different configuration space $\Phi$ and $\tilde{\Phi}$ will lead to one and the same physical configuration space $\Phi/\mathcal{G} \cong \tilde{\Phi}/\tilde{\mathcal{G}}$. Here we consider only the 'metric' configuration space for definiteness. The equivalence of the path integrals based on $g_{\mu\nu}$ and $\gamma_{\mu}$ has been argued in \cite{Gies:2013noa}. 

\section{Gaussian integrals and determinants}
\label{A2}

The one-loop formula for the generating function $W$ is given by  
 \bea \label{Z1explicit}
\mathcal{Z}  &=& \frac{1}{\Omega} \int \prod_{\bar{a}} \frac{d f^{\bar{a}}}{\sqrt{2\pi}} \sqrt{{\det}' (G^{-1})^{\alpha \beta}}  \sqrt{{\det}' C_{\alpha \beta}}   \sqrt{\det C_{\bar{a}{\bar{b}}} }\, e^{-  S[\bar{\phi}] - \frac{1}{2} f \cdot  S^{(2)}  \cdot f   }  \nonumber  \\
  &=& \frac{1}{\Omega_{KV}} \frac{\sqrt{{\det}' (G^{-1})^{\alpha\gamma}    C_{\gamma \beta}}}{ \sqrt{\det  (C^{-1})^{\bar{a}\bar{c}} \left(S^{(2)}_J \right)_{\bar{c}\bar{b}}}    }  e^{-  S_J[\bar{\phi}] }  
 \eea
 with all quantities evaluated on an Einstein space with Ricci curvature \eq{curvature} and extrinsic curvature determined by \eq{Kij} and \eq{KJ} in case of a boundary.
 It is clear that $\mathcal{Z}$ is reparameterisation invariant since it transforms as a scalar on configuration space $\Phi$. 
To compute this integral we can pick any field parameterisation and then from there determine a decomposition \eq{decompose}
which satisfies \eq{decompmetric} after a shift \eq{shift}.

Taking the field to be given by the metric tensor $\phi^A = g_{\mu\nu}$ the metric $C$ on $\Phi$ is of the DeWitt form
\beq
C_{ab} = C^{\mu\nu,\rho \sigma} \delta(x-y) = \frac{\mu^2}{32 \pi G} \sqrt{g}\left( \frac{1}{2} (g^{\mu\rho}g^{\nu\sigma} + g^{\mu\sigma}g^{\nu\rho}) -  \frac{1}{2} g^{\mu\nu} g^{\rho\sigma} \right)\delta(x-y)
\eeq
For a gauge parameter $\xi^{\alpha}= \epsilon^\mu(x)$ the corresponding metric is given by \eq{Gab}.
Proceeding as outlined in  Appendix~\ref{A1} we can first decompose $\delta g_{\mu\nu} $ into the gauge modes and the gauge invariant fields:
\beq \label{decomp}
\delta g_{\mu\nu} = h_{\mu\nu}^{\rm TT}  + \frac{1}{D} g_{\mu\nu} s   + \nabla_\mu \epsilon'_{\nu} + \nabla_\nu \eps'_\mu
\eeq
where  $h_{\mu\nu}^{\rm TT}$ is transverse and traceless. The Killing vectors are removed from $\eps'_\mu$ since these are the zero modes of $L^a_{\alpha}$. Furthermore if the background involves modes 
\beq
\nabla_{\mu} \nabla_{\nu} s = \frac{1}{D} g_{\mu\nu} \nabla^2 s
\eeq     
which, satisfy the eigen-problem $-\nabla^2 s = \frac{R}{D-1} s$ these must be removed from the spectrum of $s$.
This follows since $g_{\mu\nu} s \propto \nabla_\mu \eps^{\rm CKV}_{\nu} + \nabla_\nu \eps^{\rm CKV}_{\mu}$  where  $\eps_{\mu}^{\rm CKV}$ is a conformal Killing vector which is not a Killing vector (CKV) and is included in the spectrum of $\eps_{\mu}$.
The line element is given by
\beq
C_{ab} \delta \phi^a \delta \phi^b =  \frac{\mu^2}{32 \pi G}  \int d^Dx \sqrt{g}  \left(h_{\mu\nu}^{TT} h^{TT\mu\nu} - \frac{D-2}{2D} s^2  + 2 \eps'_{\mu} \Delta_1 \eps'^{\mu}  - 4  \frac{D-2}{2D} \nabla_{\mu} \eps'^\mu   s \right)
\eeq
where we have exploited the boundary conditions  \eq{BRST_bcs}  to integrate by parts finding that all boundary terms vanish. Since the metric $C_{ab}$ is not diagonal in these coordinates we makes the shift
\beq
s \to s - 2 \nabla_{\mu} \eps''^{\mu}
\eeq
to calculate the determinant where the double prime indicates that we do not include the CKV's or KV's in the transformation.
We then have
\beq
C_{ab} \delta \phi^a \delta \phi^b =  \int d^Dx \frac{\mu^2}{32 \pi G_0} \sqrt{g} \left(h_{\mu\nu}^{TT} h^{TT\mu\nu} - \frac{D-2}{2D} s^2     + 2 \eps''^{\mu} \tilde{\Delta}_1 \eps''_{\mu}  + 2 \sum_{\rm CKV} \eps_{\mu} \left(\frac{1}{D-1} -  \frac{2}{D} \right) R \, \eps^{\mu} \right) \nonumber
\eeq
where we have included explicitly the contribution from the CKV's and introduced the differential operator 
\beq
\tilde{\Delta}_1 \eps_{\mu} = \Delta_1  \eps_{\mu}  - \frac{D-2}{D} \nabla_{\mu} \nabla^{\nu} \epsilon_{\nu} \,.
\eeq
For the line element on the space of diffeomorphisms \eq{Gab} we have
\beq
\sqrt{{\det}' (G^{-1})^{\alpha\gamma}    C_{\gamma \beta}} = \sqrt{ \left( \frac{R/\mu^2}{D-1} - 2\frac{R/\mu^2}{D}     \right)^{N_{\rm CKV}} {\rm det}'' \tilde{\Delta}_1/\mu^2 } \,.
\eeq
We then note that the spectrum of $\tilde{\Delta}_1$ may be decomposed into transverse and longitudinal modes such that we obtain
\beq \label{Gaugedets}
\sqrt{{\det}' (G^{-1})^{\alpha\gamma}    C_{\gamma \beta}} = \sqrt{ \left( \frac{R/\mu^2}{D-1} - 2\frac{R/\mu^2}{D}     \right)^{N_{\rm CKV}} {\rm det}'_{1T} [\Delta_{1}/\mu^2]   {\rm det}'' \left[ \frac{2(D-1)}{D}  \Delta_{0}/\mu^2 \right]}
\eeq
where $1T$ indicates that is the operator $\Delta_1$  acts on transverse vectors and
\beq \label{Delta0opertor}
\Delta_{0} = -\nabla^2 + \frac{R}{D-1}
\eeq 
acts on scalars with the double prime indicating that the zero modes and constant mode should be removed from the determinant.

Now we compute the gaussian integral over the gauge invariant fields $f^{\bar{a}} = \{ h_{\mu\nu}^T,s \}$.
Taking the second variation of the action we have
\beq \label{delta2S}
\delta^2 S = \int d^dx \sqrt{g}  \left( - \frac{1}{16 \pi G} F^\mu(h) F_{\mu}(h)    + \frac{1}{32 \pi G} \bar{h}^{\mu\nu} \Delta_2 h_{\mu\nu} \right)
\eeq
where $h_{\mu\nu} = \delta g_{\mu\nu}$, $\bar{h}_{\mu\nu} = h_{\mu\nu} - \frac{1}{2} g_{\mu\nu} h^\lambda_\lambda$ and $F_{\lambda}(h) = g^{\mu\nu} \nabla_{\mu} \bar{h}_{\nu\lambda}$.
Inserting \eq{decomp} one readily finds the hessians for the gauge invariant fields
\beq \label{hTThess}
h^{TT} \cdot S^{(2)}_{h^{TT} h^{TT}} \cdot h^{TT}   = \frac{1}{32 \pi G}  \int d^Dx \sqrt{g} \,   h_{\mu\nu}^{TT} \Delta_2  h^{\mu\nu TT}
\eeq
\beq \label{shess}
s \cdot S^{(2)}_{ss} \cdot s   = - \frac{1}{32 \pi G}   \frac{(D-2)(D-1)}{D^2}  \int d^Dx \sqrt{g} \, s \Delta_0 s
\eeq
while $\eps_{\mu}$ components of the hessian are zero. Here we see that the hessian for $s$ has the wrong sign for all modes where $\Delta_0$ is positive which corresponds to all modes apart from the constant mode $s_0$ when $R>0$.
To ensure that the Wick rotation gives a well defined Euclidean theory. This can be achieved by canonically normalising the scalars
\beq
s \to \sqrt{ -\frac{32 \pi G D^2}{(D-2)(D-1)}}  s
\eeq
 Such that hessian for $s$ is then given by
\beq
s \cdot S^{(2)}_{TT} \cdot s   =    \int d^Dx \sqrt{g} \, s \left( -\nabla^2 - \frac{R}{D-1} \right)  s
\eeq
additionally we canonically normalise $h_{\mu\nu}^{TT}$ via
\beq
h_{\mu\nu}^{TT} \to \sqrt{ 32 \pi G} h_{\mu\nu}^{TT} \,.
\eeq
Then the metric on the gauge invariant field space becomes
\beq
C_{\bar{a}\bar{b}}  f^{\bar{a}} f^{\bar{b}} =  \int d^Dx \mu^2 \sqrt{g} \left(h_{\mu\nu}^{TT} h^{TT\mu\nu} +  \frac{D}{2(D-1)}  s^2  \right)
\eeq
and we must remember that the constant mode of $s$ must be Wick rotated back (since the gaussian integral originally had the correct sign). 
We then have
\beq
\sqrt{\det | (C^{-1})^{\bar{a}\bar{c}} \left(S^{(2)} \right)_{\bar{c}\bar{b}}|}  = \sqrt{  {\det}_{2T^2} [\Delta_2/\mu^2] {\rm det}' \left| \frac{2(D-1)}{D} \Delta_0/\mu^2 \right|} \,.
\eeq
Comparing this expression with \eq{Gaugedets} we observe that integral over the scalar modes cancels with the the determinant from factoring out the longitudinal diffeomorphisms apart from the CKVs and  the constant mode.
Here $2T^2$ means the determinate is over transverse-traceless modes.

To check that the final result will not depend on the choice of field parameterisation $\phi$ (or equivalently the coefficients \eq{Tcoefficients}) we note that terms involving $\delta^2 g_{\mu\nu}$ are not present since we expand around the saddle point and any dependence on $\mathcal{T}_{A\mu\nu}$ cancels between the determinates in \eq{Z1explicit}. 
We therefore have:
\bea \label{W1explicit1}
-\log \mathcal{Z} &&= S[\bar{\phi}] + \frac{1}{2} \Tr_{2 T^2} \log \Delta_2/\mu^2 + \frac{1}{2} \log \frac{2}{D} |R|/\mu^2  - \frac{1}{2} N_{\rm CKV} \log \left((1/(-1 + D) - 2/D) R/\mu^2 \right) \nonumber \\
&&  - \frac{1}{2}  \Tr_{1T}' \log \Delta_1/\mu^2 - \log \Omega  
\eea
independently of the gauge or field parameterisation.
Finally, using the relations between traces of constrained fields and unconstrained fields on an arbitrary Einstein space (see e.g appendix B of \cite{Falls:2016msz}):
\beq
 \Tr_{2T^2} f (\Delta_2)   = \Tr_{2}  f(\Delta_2)  -\Tr'_{1}  f(\Delta_1)  -   \Tr_0 f(-\nabla^2 - \frac{2}{D} R)   +   N_{\rm CKV} f \left((1/(-1 + D) - 2/D) R \right)   \,,
\eeq
\beq
 \Tr_{1T}' f (\Delta_1)   = \Tr_{1}'  f(\Delta_1)   - \Tr_0 f(-\nabla^2 - \frac{2}{D} R) +  f(- \frac{2}{D} R) \,,
\eeq
we can then arrive at \eq{unregulated} where the traces are for unconstrained symmetric tensors and vectors \footnote{Here we have neglected a constant imaginary part which is needed to correct the contribution of the zero mode ensuring $W$ is real for $R>0$}. It is straightforward to check that using the gauge fixing \eq{gaugefixing} with $\alpha =1$ gives the same result since the gauge fixing action cancels the first term in \eq{delta2S} and the corresponding Faddeev-Popov determinant gives the vector trace. Upon replacing the $C_{ab}$ and $G_{\alpha\beta}$ with the regulated forms \eq{CLambda} and \eq{GLambda} we then obtain the traces \eq{W1reg} which are free from divergencies.

\section{The gauge invariant hessian for general schemes near two dimensions}
\label{Hessians_in_two_dimensions}
Close to two dimensions we are interested in matter with an interaction
\beq
\mathcal{O} = \int d^Dx \sqrt{g} \mathcal{L}(\psi) 
\eeq
by writing going to dimensionless matter fields rescaled by the determinant of $g_{\mu\nu}$ to the appropriate power. Then the interaction becomes
\beq
\mathcal{O}[\phi] = \int d^Dx \sqrt{g}^{-d_{0}/D}(\phi) \mathcal{L}(\hat{\psi}(\phi)) 
\eeq
and the Kinetic terms become invariant under conformal transformations of the metric holding $\hat{\psi}$ fixed.
Then the set of conformal gauges
\beq \label{conformal_gauge}
g_{\mu\nu} = f(\sigma) \hat{g}_{\mu\nu}
\eeq
with the determinate of $\hat{g}_{\mu\nu}$ fixed, become useful since the kinetic terms are then in dependent of $\sigma$. 
We then replace in the last section $\lambda \mathcal{V} \to \g \mathcal{O}$ and repeat the analysis.
The calculation is simplest in the conformal gauges however since we only need the on-shell hessian to find the divergencies of $\mathcal{Z}$ all terms that depend on this choice 
vanish once we use the equations of motion. This results the operator $\Delta_0$ in \eq{shess} by being replaced by
\beq \label{Delta0general}
\Delta_0 \to -\nabla^2 - \frac{d_{0}}{2}{R}
\eeq
for $D \to 2$ and produce a term which mixes between gravity and matter which vanishes as $D \to 2$. This agrees with \eq{Delta0opertor} in the case $d_0=-2$.
Since the kinetic term for the matter fields is conformally invariant there is no mixing between $\hat{\psi}$ and $\sigma$ from this term. From $\mathcal{O}$ there is 
a component of the hessian that mixes $\sigma$ and $\hat{\psi}$ however this term only contributes to irrelevant power law divergencies and not the universal beta functions. 
It follows that \eq{Delta0general} is the only significant difference between the the on-shell Hessians for the case $ \mathcal{V} = \mathcal{O}$ and the general case.  As such we arrive at
 the flow equation where the one-loop coarse graining contribution is given by \eq{F_general}.

\section{Volume of the stability group $\mathcal{H}$}
\label{A3}

 Non-perturbatively the the volume $\Omega$ of the stability group $\mathcal{H}$ takes the form \cite{Volkov:2000ih}
 \beq \label{Omega}
 \Omega(\mu) = \prod_{\ell=1}^{N_{KV}}   \int d \mathcal{M}(\epsilon_{\ell})  \frac{\mu^2}{\sqrt{\pi}} ||k_\ell|| 
 \eeq
 involving the Haar measure on $\mathcal{H}$
 where $||k_\ell||\equiv \sqrt{ \langle k_\ell |k_\ell \rangle }$  is the square root of the norm
 \beq
 \langle k_\ell |k_{\ell'} \rangle = \frac{1}{32 \pi G} \int d^Dx \sqrt{g} \,  k_{\ell}^\mu k_{\ell'}^\nu g_{\mu\nu}
 \eeq
 $k_{\ell}^\mu = \frac{\partial x^\mu}{ \partial \epsilon_\ell} $ are Killing vectors where  $\langle k_\ell |k_{\ell'} \rangle = 0$ for $\ell \neq \ell'$  where we have decomposed $\epsilon^\mu_{KV} = \sum_{\ell=1}^{N_{KV}}  \epsilon_{\ell} k^\mu_{\ell}$.
 The volume $\Omega$ has been calculated explicitly for both $S^4$ and $S^2 \times S^2$ space-times in \cite{Volkov:2000ih} (denoted there by $\Omega_1$). 
 We note that the proper-time regularisation replaces $\mu^2$ with $ \Lambda^2   e^{-\gamma_E}$  in \eq{Omega} such that
 \beq \label{LambdaOmega}
 \Lambda \partial_\Lambda  \Omega = 2 N_{KV} \Omega \,.
 \eeq
 which is important to obtain background independent beta functions.

\newpage

\section{Boundaries}
\label{boundaryApp}
For a manifold of dimension $D$ the boundary is located at $f(x) = 0$ and has coordinates $y^i$ giving rise to tangent vectors $e^\mu_i = \frac{\partial x^{\mu}}{\partial y^i}$.
The normal vector is defined by
\beq
n_{\rho} = \frac{f_{,\rho}}{\sqrt{g^{\mu\nu} \partial_\mu f \partial_\nu f}} \,,
\eeq
along with the condition $n^\nu \nabla_\nu n^\mu = 0$. The induced metric and extrinsic curvature  are defined by
 \beq
 \ga_{ij} = e^\mu_i e^\nu_j g_{\mu\nu}\,,  \,\,\,\,\,\,\,\,  K_{ij} = e^\mu_i e^\nu_j \nabla_{\nu} n_{\mu}
 \eeq
and $K =  \ga^{ij} K_{ij}$. Denoting covariant derivatives in the boundary by $|$ and a normal derivative by a dot one has the following useful identities for vectors
\bea
\nabla_j \eps_{i} &=& \eps_{i|j} + K_{ij} \eps_n \\
\nabla_j \eps_{n} &=& \eps_{n|j} -  K_j\,^i \eps_i\\
\nabla_n \eps_{i} &=& \dot{\eps}_i \\
\nabla_n \eps_{n} &=&  \dot{\eps}_n
\eea
and for symmetric tensors
\bea
\nabla_k h_{ij}&=&  h_{ij|k}+2 h_{n (i}  K_{j)k} \\
\nabla_j h_{in} &=& h_{i|j} + h   K_{ij} - h_{i a}   K^a\,_{j} \\
\nabla_k h_{nn}&=& h_{n|k} - 2 h_{na} K^a_k\\
\nabla_n h_{ij} &=&   \dot{h}_{ij}\\
\nabla_n h_{in} &=& \dot{h}_{ni}\\
\nabla_n h_{nn} &=& \dot{h}_{nn}
\eea
To show that the third boundary condition \eq{BRST_bcs} is diffeomorphism invariant one must use that $n^{\mu} \Delta_1 \eps_{\mu} \propto \eps_{n} = 0$ which follows from
expanding $\eps_{\mu}$ in the eigen-basis corresponding to $\Delta_1$.

Defining $h^{\eps}_{\mu\nu} = \nabla_\mu\eps_{\nu} + \nabla_\nu \eps_\nu$ one can show that
\beq \label{Intbyparts}
 \int d^dx \sqrt{g} \,\bar{h}^{\epsilon}_{\mu\nu} \Delta_2 h^{\mu\nu}   =  \int d^dx \sqrt{g} \, \bar{h}^{\mu\nu} \Delta_2  h^{\eps}_{\mu\nu}  \,.
\eeq
where all boundary terms cancel after integrating by parts and using the boundary conditions \eq{BRST_bcs}. Interestingly this cancelation is related to the tensor structure of the field space metric \eq{deWittmetric1}.  
The second variation of the action \eq{Sb} subject to the boundary conditions is given by \eq{delta2S}. To show that the $\eps_{\mu}$ components of the hessian are zero and the hessians of the gauge invariant fields are given by \eq{hTThess}  and \eq{shess} is straightword.
To do so one  makes use of \eq{Intbyparts},
\bea
F_{\nu}(h^\eps) &=& - \Delta_1 \eps_\mu \,,  \\
 \Delta_2 h^\eps_{\mu\nu} &=& h^{\Delta_1 \epsilon}_{\mu\nu}\\
 & =& \nabla_{\mu} \Delta_1 \eps_{\nu} +  \nabla_{\nu} \Delta_1 \eps_{\mu} \\
 & =& - \nabla_{\mu} F_{\nu}(h^\eps) - \nabla_\nu F_{\mu}(h^\eps)  
\eea
 and that
\beq
n^{\mu} \Delta_1 \eps_\mu =0 \,, \,\,\,\, n^{\mu} F_{\mu}(h) = 0 \,
\eeq
both vanish on the boundary $\Sigma$.

 \end{appendix}

  \bibliography{myrefs,ASreferences}

\end{document}